\documentclass[prb,aps,twocolumn,superscriptaddress,showpacs]{revtex4}
\usepackage{amsmath}
\usepackage{amssymb}
\usepackage{amsfonts}
\usepackage[pdftex]{graphicx}		
\usepackage{array}	
\usepackage{bm}		
\usepackage{dsfont}	
\usepackage{ulem}

\begin{document}

\title{Strong coupling behavior of the neutron resonance mode in
  unconventional superconductors}

\author{Patrik Hlobil}
\affiliation{Karlsruher Institut f\"ur Technologie, Institut f\"ur Theorie der Kondensierten
Materie, 76128 Karlsruhe, Germany}
\author{Boris Narozhny}
\affiliation{Karlsruher Institut f\"ur Technologie, Institut f\"ur Theorie der Kondensierten
Materie, 76128 Karlsruhe, Germany}
\author{J\"org Schmalian}
\affiliation{Karlsruher Institut f\"ur Technologie, Institut f\"ur Theorie der Kondensierten
Materie, 76128 Karlsruhe, Germany}
\affiliation{Karlsruher Institut f\"ur Technologie, Institut f\"ur Festk\"orperphysik, 76128
Karlsruhe, Germany}
\date{\today }

\begin{abstract}
 We analyze whether and how the neutron resonance mode in
 unconventional superconductors is affected by higher order
 corrections in the coupling between spin excitations and fermionic
 quasiparticles and find that in general such corrections cannot be
 ignored. In particular, we show that in two spatial dimensions
 ($d=2$) the corrections are of same order as the leading, one-loop contributions demonstrating that the neutron resonance mode
 in unconventional superconductors is a strong coupling
 phenomenon. The origin of this behavior lies in the quantum-critical
 nature of the low energy spin dynamics in the superconducting
 state and the feedback of the resonance mode onto the fermionic excitations. While  quantum critical fluctuations occur in any
 dimensionality $d\leqslant 3$, they can be  analyzed in a controlled
 fashion by means of the $\varepsilon$-expansion ($\varepsilon =3-d$),
 such that the leading corrections to the resonance mode position are
 small. Regardless of the strong coupling nature of the
 resonance mode we show that it emerges only if the phase of the superconducting gap
 function varies on the Fermi surface, making it a powerful tool to
 investigate the microscopic structure of the pair condensate.
\end{abstract}

\maketitle


The emergence of a resonance mode in the inelastic spin excitation
spectrum below the superconducting transition temperature has become
an important indicator for unconventional superconductivity in a range
of correlated materials. First observed
\cite{RossatMignot91,Mook1993,Fong1995,Fong1996,Bourges1996} in
YBa$_{2}$Cu$_{3}$O$_{7-\delta }$, the phenomenon occurs in other
cuprate superconductors~\cite{Fong1999,He2001,He2002}, in
heavy-electron superconductors~\cite{Petrovic2001,Stock2008,Stockert2008}, and in
iron-based materials~\cite{Christianson2008,Inosov2010}. Below
$T_{c}$, one observes essentially two effects in the inelastic neutron
spectrum: (i) the low-energy spectral weight is suppressed for
energies $\omega<2\Delta$, where $\Delta$ is the magnitude of the
superconducting gap; and (ii) a sharp peak occurs at
$\Omega_{\mathrm{res}}<2\Delta$ that is centered around a finite
momentum $\mathbf{Q}$. Usually, $\mathbf{Q}$ coincides with the
ordering vector of a nearby antiferromagnetic state. For $T=0$, the imaginary
part of the dynamic spin susceptibility at the momentum $\mathbf{Q}$
can be described as
\begin{equation}
{\rm Im} \chi _{\mathbf{Q}}\left( \omega \right) =
Z_{\mathrm{res}}\delta\left( \omega -\Omega _{\text{res}}\right) 
+ {\rm Im} \chi _{\mathbf{Q}}^{\mathrm{inc}}\left( \omega \right) ,  
\label{r1}
\end{equation}
where $Z_{\mathrm{res}}$ is the spectral weight of the resonance mode
while the imaginary part of the incoherent part
$\chi_{\mathbf{Q}}^{\mathrm{inc}}\left(\omega\right)$ vanishes for
$|\omega|<2\Delta$.

\begin{figure}
\centering
\includegraphics[width=0.43\textwidth]{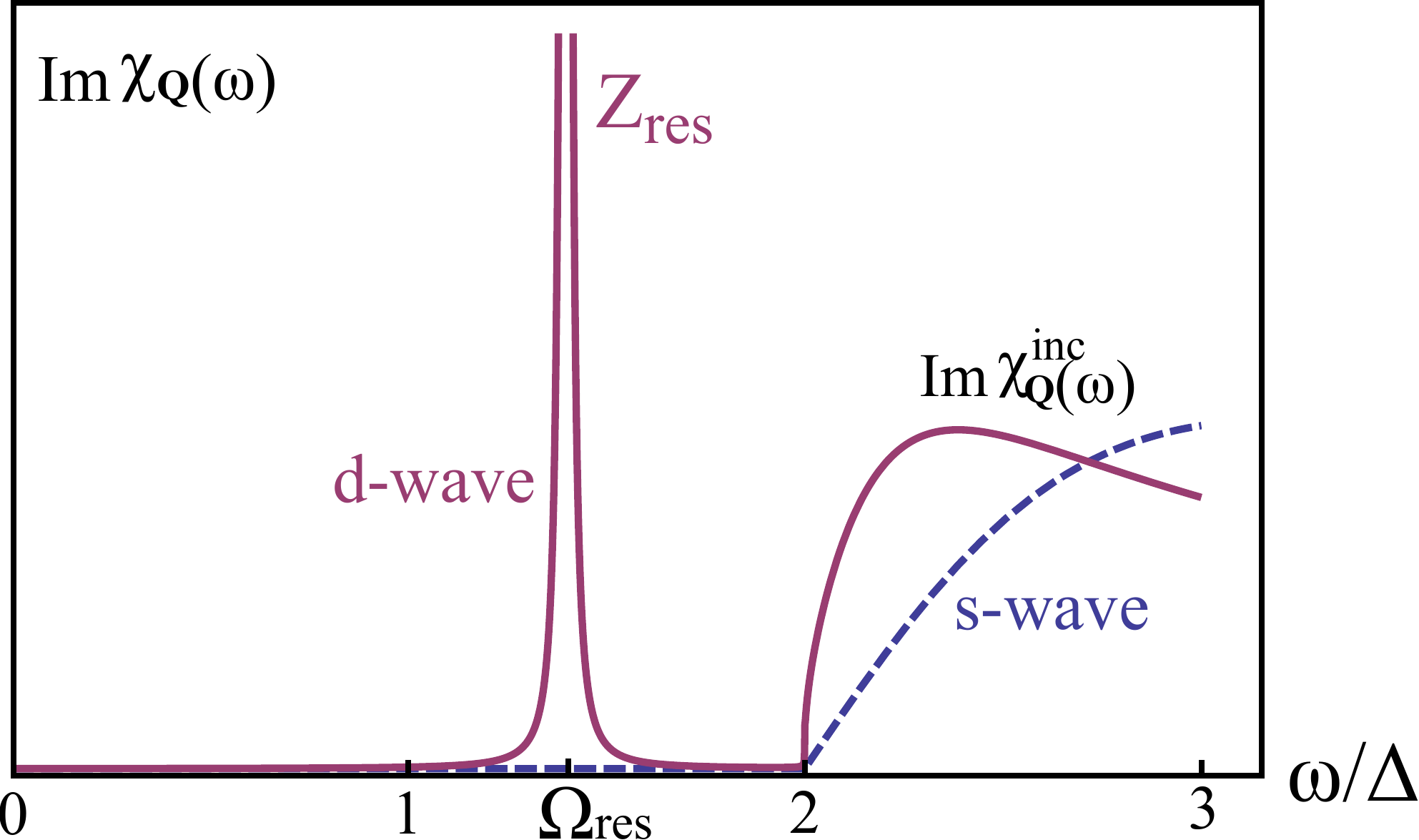}
\caption{Spin spectrum for different gap symmetries in the superconducting state. For sign-changing gap symmetry a resonance occurs at $\Omega_{\text{res}}<2 \Delta$ in the spin gap and the continuum is governed by particle-hole damping of superconducting excitations.}
\label{PI1}
\end{figure}

A promising explanation for the resonance mode that permits detailed
comparison with experiment was obtained within an one-loop
approach
\cite{Abanov2000,Eschrig2000,Abanov2001,Eschrig2002,Abanov2002,Eschrigreview,Korshunov2008,Maier2009,Eremin2005,Manske2001}.
Within this approach, collective excitations of the superconductor are
sensitive to the coherence factors of the BCS-like wave function. The coherence factors
determine scattering-matrix elements for (i) interactions
between Bogoliubov quasiparticles and (ii) interactions between
quasiparticles and the pair condensate. In the case of spin-spin
coupling (where the scattering matrix is odd under time reversal), the
latter processes leads to the emergence of the resonance mode if the phase
of the superconducting gap function $\Delta_{\mathbf{k}}$ takes
distinct phases at momenta $\mathbf{k}_{F}$ and
$\mathbf{k}_{F}+\mathbf{Q}$ (assuming that both belong to the Fermi
surface). This effect makes neutron scattering sensitive to the
internal structure of condensed pairs and allows one to identify
unconventional pairing.

Unconventional superconductivity often occurs in close proximity of
competing states with long-range order. Consequently, the concomitant
quantum criticality requires an investigation of the microscopic
structure of the superconducting state (and in particular, of the
resonance mode) that goes beyond the usual one-loop approach.  It
is well known, that itinerant systems in the vicinity of a spin-density-wave quantum-critical point are characterized by the energy
scale $\omega_{\text{sf}}\propto\xi^{-2}$ of the normal state spin
excitation spectrum~\cite{Hertz1976,Moriya1985,Millis1993} that
vanishes at the quantum critical point, where the magnetic correlation
length $\xi$ diverges. As a result, precisely at those points on the
Fermi surface that are connected by the magnetic ordering vector
(i.e. $\mathbf{k}_{F}$ and $\mathbf{k}_{F}+\mathbf{Q}$) the
quasiparticle lifetime for energies above $\omega _{\text{sf}}$
deviates from the standard Fermi-liquid result~\cite{Abanov2003,Metlitski2010}. In two- or three-dimensional
systems, these sets of points are referred to as hot spots or hot
lines of the Fermi surface, respectively. So far, it is unclear
whether or not quantum-critical fluctuations that are relevant at
higher energies $\omega >\omega _{\text{sf}}$ and contribute to the
incoherent contribution ${\rm Im} \chi
_{\mathbf{Q}}^{\mathrm{inc}}\left( \omega \right) $ in Eq.~(\ref{r1})
lead to any feedback on the spectral features of the resonance
mode. For example, if higher order vertex corrections to the dynamic
spin susceptibility are governed by excitations with energies smaller
than $\omega_{\text{sf}}$ (and thus behave similar to Fermi-liquid
quasiparticles), then the weak coupling picture is expected to be
robust. On the other hand, if such virtual excitations are quantum
critical, i.e. have typical energies larger than $\omega_{\text{sf}}$,
the analysis becomes more subtle \cite{me,Onufrieva}.

Another open issue is related to the sensitivity of the
resonance mode with respect to the variation of the phase of the superconducting order parameter on the Fermi surface. Whether or not this is the case if one takes into account higher orders in perturbation theory needs to be explored. The relevance
of strong coupling behavior for the  resonance
mode is also suggested by the observation of a nearly universal ratio
of $\Omega _{\mathrm{res}}$ and $\Delta $ in a wide range of
systems~\cite{Greven2009}, which one would not expect from
weak coupling theory.

In this paper we evaluate self-energy and vertex corrections to the
dynamic spin susceptibility in the superconducting state and determine
higher order corrections to the neutron-resonance mode. For $d=2$, we find that both, self-energy and vertex corrections, cause significant changes in the resonance and cannot be ignored, except for very weak coupling strength.  Near a magnetic quantum-critical point these corrections are of same
order as the leading one-loop result, revealing that the
resonance mode is a strong coupling phenomenon. Self-energy corrections are primarily caused by singularities in the fermionic spectrum that were caused by the resonance mode in the first place. In contrast, vertex corrections are dominated by quantum-critical
fluctuations contributing to ${\rm Im}\chi_{\mathbf{Q}}^{\mathrm{inc}}\left(\omega\right)$, due to the fact that virtual processes lead to the
emergence of the resonance mode.   In order to develop a controlled theory of the resonance mode, we perform an $\varepsilon$-expansion around the upper critical dimension $d_{\text{uc}}=3$, that reveals how
quantum-critical fluctuations affect the dynamic spin susceptibility
as function of the dimensionality of the system. These results
demonstrate that the theory of
Refs.~\onlinecite{Abanov2000,Eschrig2000,Abanov2001,Eschrig2002,Abanov2002,Eschrigreview,Korshunov2008,Maier2009,Eremin2005,Manske2001}
is applicable for three-dimensional superconductors including
moderately anisotropic  materials. On the other hand, for
$d=2$ the neutron resonance mode is a strong coupling phenomenon and
our results show that no controlled theory for the effect exists so
far. Finally, we demonstrate that  higher-order vertex corrections
only lead to a resonance mode if the phases of the gap at $\Delta
_{\mathbf{k}_{F}}$ and $\Delta _{\mathbf{k}_{F}+\mathbf{Q}}$ are
distinct.

\section{The spin fermion model}

Consider an unconventional superconductor in the vicinity of a spin
density wave instability. Low-energy spin excitations of the system
(i.e. paramagnons) can be described~\cite{Abanov2003,Metlitski2010} in
terms of a spin-1 boson $\mathbf{S}_{\mathbf{q}}$ that is
characterized by the dynamic spin susceptibility
\begin{equation}
\chi_{\mathbf{q}}\left( \omega \right) =\frac{1}{r_{0}+c_{s}
\left(\mathbf{q}-\mathbf{Q}\right)^{2}-\Pi_{\mathbf{q}}\left(\omega\right)},
\label{spinsus}
\end{equation}
where ${\bf q}$ and $\omega$ are the wave-vector and frequency,
$\mathbf{Q}$ is the antiferromagnetic ordering vector and $r_{0}$
determines the distance to the instability. The spin dynamics is described by the
self-energy $\Pi_{\mathbf{q}}(\omega)$. Hereafter,
$\chi_{\mathbf{q}}(\omega)$ refers to the retarded susceptibility,
while $\chi_{\mathbf{q}}(i\omega _{n})$ is used for the corresponding
Matsubara function. Similar notations are used below for fermionic
Green's functions and self-energies. The spin dynamics, encoded in
$\Pi_{\mathbf{q}}(\omega)$, is a consequence of coupling of the
collective spin degrees of freedom ${\bf S}_{\bf q}$ to low-energy
particle-hole excitations. At low energies and in the normal state,
the dominant contribution to the imaginary part of $\Pi_{\mathbf{q}}(\omega)$ for $\mathbf q \approx \mathbf Q$ comes from the
fermionic quasiparticles in the vicinity of the hot spots (or hot
lines) on the Fermi surface (defined by the relation
$\varepsilon_{\mathbf{k}_F+\mathbf{Q}}=\varepsilon _{\mathbf{k}_F}$,
where $\varepsilon _{\mathbf{k}}$ is the bare fermionic
single-particle dispersion measured relative to the Fermi energy). In
this paper, we consider commensurate magnetic order, where
$2\mathbf{Q}$ is equal to a reciprocal lattice vector,
i.e. $\varepsilon _{\mathbf{k}+2\mathbf{Q}}=\varepsilon
_{\mathbf{k}}$.

Let $\psi_{\mathbf{k}\alpha}^{\dagger}$ be the creation operator of a
fermion with the momentum $\mathbf{k}$ and spin index $\alpha$. Coupling between spin fluctuations and fermionic quasiparticles is
described by the following term in the
Hamiltonian~\cite{Abanov2003,Metlitski2010}
\begin{equation}
\mathcal{H}_{int}=g\int d^dx \,   \mathbf S \cdot (\psi_\alpha^{\dagger}
\bm \sigma_{\alpha\beta}\psi_\beta),
\end{equation}
where operators are given in real space and $\bm\sigma$ is the vector
of the Pauli matrices. A microscopic derivation of this model is
possible in the limit of weak electron-electron interactions and may
be based upon a partial resummation of diagrams in the particle-hole
spin-triplet channel. In this case, it is usually not permissible to
approach the regime in the close proximity of the magnetic critical
point, which for generic Fermi surface shape requires a threshold
strength of the interaction. However, we may consider this
spin-fermion model as a phenomenological theory of low-energy
quasiparticles coupled to spin fluctuations that is valid only at
energies small compared to the initial electron bandwidth.  At low
energies, fermions near the hot spots determine the spin dynamics and
are crucial for the spin-fluctuation-induced pairing state. In this
case the electronic spectrum near the hot spots may be linearized,
$\varepsilon_{\mathbf{k}}=\mathbf{v}_{\mathbf{k}}\cdot\mathbf{k}$
(where $\mathbf{v}_{\mathbf{k}}$ is the quasiparticle velocity at
momentum $\mathbf{k}$). In what follows, we assume that $g$ is small
compared to the corresponding fermionic scales, which implies
smallness of the dimensionless parameter $\gamma$
\begin{equation}
\frac{\gamma}{N} =\frac{g^{2}}{4\pi v_{\parallel }v_{\perp }}k_{F}^{d-2} \ll 1.
\label{gamma}
\end{equation}
Here $v_\|$ and $v_\perp$ are the projections of
$\mathbf{v}_{\mathbf{k}}$ at the hot spot onto directions that are
parallel and perpendicular to $\mathbf{Q}$, respectively, and $N$ is
the number of hot spots (lines), or generally the number of fermion
flavors that couple to the spin excitations. In what follows, we
assume $v_{\parallel }=v_{\perp } = v_F/\sqrt{2}$ and use
$v_{F}=\left\vert \mathbf{v}\right\vert $, see
Fig.~\ref{hotspot}.

The spin-fermion model can be described by an effective
action that in the superconducting state takes the form
\begin{eqnarray}
S &=&-\frac{1}{2}\int_k\Psi_k^\dagger \widehat G_{0,k}^{-1}\Psi_k
+\frac{1}{2}\int_q\chi_{0,q}^{-1}\mathbf{S}_q\cdot \mathbf{S}_{-q}
\nonumber\\
&&
\nonumber\\
&&
+g\int_{k,k^{\prime }} (\Psi_k^\dagger \hat{\bm \alpha} \Psi_k) \cdot 
\mathbf S_{k-k^{\prime }},  
\label{sfsc}
\end{eqnarray}
where 
\[
\Psi_k=\left( 
\begin{matrix}
\psi_{k\uparrow } & \psi_{k\downarrow } & \psi_{-k\uparrow }^\dagger & 
\psi_{-k\downarrow }^\dagger
\end{matrix}
\right)^T
\] 
is the extended Gor'kov-Nambu-spinor. Here, we use the following notations
\begin{equation*}
\hat\alpha ^i=\left( 
\begin{matrix}
\sigma^i & 0 \cr 
0 & \sigma^y\sigma^i\sigma^y
\end{matrix}
\right) \text{ and } \hat \beta =\left( 
\begin{matrix}
\mathds 1_2 & 0 \\ 
0 & \mathds -1_2%
\end{matrix}%
\right) .
\end{equation*}
The matrices $\sigma^i$  are the usual Pauli matrices and  in Eq.~(\ref{sfsc}) we
combine the Matsubara frequencies and momenta into
$k=(\mathbf{k},i\omega_n)$ and use the short-hand notation 
\[
\int_k\ldots=T\sum_n\int\frac{d^dk}{(2\pi)^d}\ldots  \, .
\]
In the basis of the extended spinor $\Psi_k$, the bare
fermion propagator is given by
\begin{equation}
\widehat G_{0,k}^{-1}=i\omega_n \hat {\mathds 1}-\varepsilon_\mathbf{k}\hat\beta, 
\end{equation}
The corresponding self-energy matrix in the superconducting state can be written as
\begin{equation}
\widehat\Sigma_k=i\omega_n\left(1-Z_k\right) 
 \hat {\mathds 1}+\delta \epsilon_k \, \hat \beta  +\Phi_k\,  \hat \alpha^\Delta + \Phi_k^* \hat \alpha^{\Delta^*},    \label{Sigmadef}
\end{equation}
where we defined the matrices
\begin{equation}
\hat \alpha ^{\Delta }=\left( 
\begin{matrix}
0 & i\sigma ^{y} \\ 
0 & 0%
\end{matrix}%
\right)  \text{ and } \hat \alpha ^{\Delta^*}=\left( 
\begin{matrix}
0 & 0 \\ 
-i\sigma ^{y} & 0%
\end{matrix}%
\right) .
\end{equation}
Using this definitions the dressed Green's function can be expressed as
\begin{align}
\widehat{\cal G}_{k}^{-1} =  \widehat G_{0,k}^{-1} - \widehat\Sigma_k .
 \end{align} 
Explicitly, we obtain for the matrix Green's function:
\begin{align}
\widehat{\cal G}_{k} &= \frac{i \omega_n Z_k \hat \alpha^0+(\epsilon_k+\delta \epsilon_k) \hat \beta + \Phi_k \hat \alpha^\Delta}{(i \omega_n Z_k)^2 - (\epsilon_k+\delta \epsilon_k)^2-\Phi_k^2}   \nonumber \\
&= \left( \begin{matrix}
{\cal G}_{k}^{(p)} & 0 & 0 & {\cal F}_{k} \\
0 & {\cal G}_{k}^{(p)} & -{\cal F}_{k}  & 0 \\
0 & -{\cal F}_{k}^*  & {\cal G}_{k}^{(h)} & 0 \\
{\cal F}_{k}^*  & 0 & 0 &  {\cal G}_{k}^{(h)}
\end{matrix}   \right)  . \label{propagator}
\end{align}
The resulting gap function $\Delta_k=\Phi_k/Z_k$ will, as usual, be
determined from the solution of the corresponding self-consistency
equations.

\subsection{Normal-state behavior}   \label{normalstatesec}

In the normal state spin fluctuations can decay into gapless
electron-hole excitations which leads to overdamped spin dynamics in
agreement with observations obtained in various neutron scattering
experiments~\cite{RossatMignon1991,Inosov2010}.
The corresponding dynamic susceptibility
\begin{equation}
\chi _{\mathbf{q}}(\omega) = 
\frac{1}
{r+c_{s}\left(\mathbf{q}-\mathbf{Q}\right)^2+i\gamma\omega},
\end{equation}
where $\gamma$ is given by Eq.~(\ref{gamma}), can be obtained
by evaluating the bosonic self-energy~\cite{Abanov2000}
\begin{equation}
\Pi_q=-2g^2\int_k {\cal G}_{0,k}^{(p)} {\cal G}_{0,k+q}^{(p)}.    \label{pi12}
\end{equation}
\begin{figure}
\centering
\includegraphics[width=0.4\textwidth]{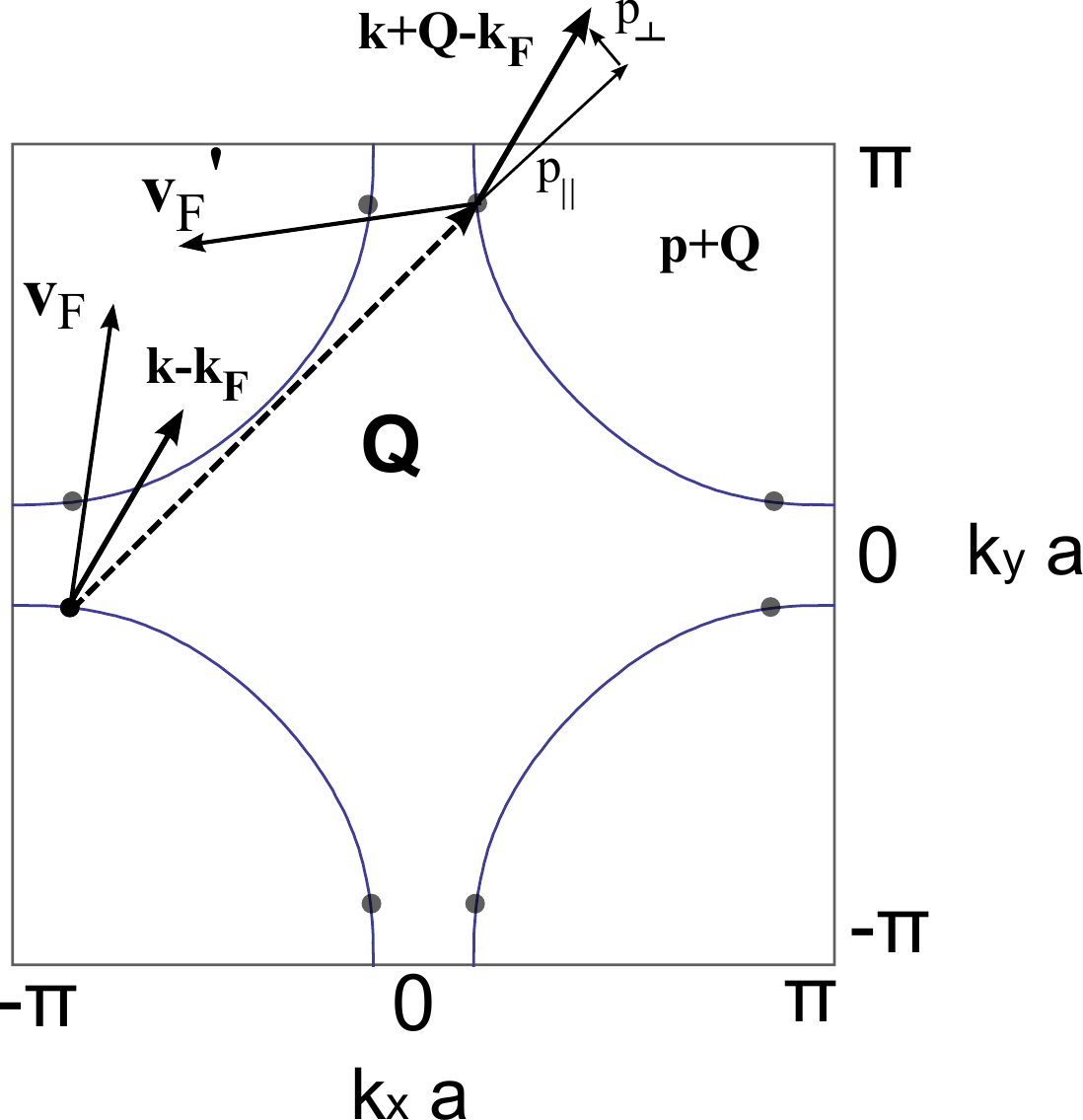}
\caption{Parametrization of the linearization of a hole-like Fermi surface with corresponding AF vector $\mathbf Q=(\pi,\pi)$. For the visualization of the Fermi surface experimental fits to a tight binding model of Bi-2212 were used~\cite{Kordyuk2003}.}
\label{hotspot}
\end{figure}
In order to calculate the one-loop diagrams it is convenient to linearize the spectrum around the hot spots $\mathbf k= \mathbf k_F + \mathbf p$ with $|\mathbf p|\ll k_F$, which dominate the integrals. As can be seen in Figure~\ref{hotspot} we can linearize $\varepsilon_{\mathbf k} \approx \frac{\mathbf k_F \cdot \mathbf p}{m} =\mathbf{v}_F \cdot \mathbf p $, where $\mathbf v_F$ is the Fermi velocity at the corresponding hot spot. Each of the $N$ hot spots contribute equally, which allows us to focus on one of them.  Along the same lines  we can also linearize the connected hot spot at $\mathbf k_F'=\mathbf k_F+\mathbf Q$ via
\begin{align}
\begin{split}
\varepsilon_{\mathbf k \approx \mathbf k_F}  &= \mathbf{v}_F \cdot \mathbf p = v_\perp p_\perp + v_\parallel p_\parallel ,  \\
\varepsilon_{\mathbf k+\mathbf Q \approx \mathbf k_F'}  &= \mathbf{v}_F' \cdot \mathbf p = v_\perp p_\perp - v_\parallel p_\parallel  .
\end{split}
\end{align}
The velocities $v_\perp$ and $v_\parallel$ are the perpendicular and parallel projections of $\mathbf v_F$ on $\mathbf Q$. Introducing  new integration variables $\epsilon=v_\perp p_\perp+  v_\parallel p_\parallel ,\epsilon'=v_\perp p_\perp-  v_\parallel p_\parallel$ it is now possible to approximate
\begin{align}
\frac{1}{L^2}& \sum_{\mathbf k} f(\varepsilon_{\mathbf k},\varepsilon_{\mathbf k+ \mathbf Q},\Delta_{\mathbf k},\Delta_{\mathbf k+\mathbf Q})   \nonumber \\
&= \frac{N}{8 \pi^2 v_\perp v_\parallel} \int d\epsilon \, d\epsilon' f(\epsilon,\epsilon',\Delta_{\mathbf k_F},\pm \Delta_{\mathbf k_F})  \, . \label{appendix1.5}
 \end{align} 
The $\pm$ signs refer to  different  gap symmetries; we consider $\Delta_{\mathbf k_F}= \pm \Delta_{\mathbf k_F+\mathbf Q}$ to be constant around the hot spots. In order to simplify our calculations we will set $v_\parallel=v_\perp = v_F/\sqrt{2}$ in future calculation, which is a suitable approximation for many known unconventional superconductors like Bi-2212 (compare with Fig.~\ref{hotspot}) .

Under the assumption that we can neglect the momentum dependence of the self-energy near the hot spots, the self-energy (\ref{pi12}) yields 
\[
\Pi_{\mathbf{Q}}(\omega)=\Pi_{\mathbf{Q}}(0)-i\gamma\omega.
\]
The static contribution $\Pi_{\mathbf{Q}}(0)$ renormalizes the bare
``mass'' $r_0\rightarrow r=r_0-\Pi_{\mathbf{Q}}^{R}(0)$ and determines
the correlation length $\xi $ via $r=c_s\xi^{-2}$.

In two dimensions ($d=2$), coupling of normal-state fermionic
quasiparticles with overdamped spin fluctuations leads to
renormalization of the fermionic spectrum. Already at one-loop level,
one finds non-trivial behavior of the fermionic self-energy at the
hot-spots~\cite{Abanov2003}:
\begin{align}
\Sigma_{\mathbf{k_F}}^{(p)} \left( i\omega _{n}\right)  
&=&
-i\frac{3g^{2}\mathrm{sign}\left(\omega_n\right)}{2\pi v_F\sqrt{c_s\gamma}}
\left(\sqrt{\omega_{\text{sf}}+|\omega_n|}-\sqrt{\omega_{\text{sf}}} \, \right) .
\label{nsse}
\end{align}
Here the frequency $\omega_{\text{sf}}=r/\gamma$ plays the role of the
crossover scale. Indeed, for energies below $\omega_{\text{sf}}$ the
self-energy (\ref{nsse}) may be approximated by the Fermi-liquid-like
expression $\Sigma(i\omega_n)=-i\omega_n\lambda$ with the dimensionless
coupling constant $\lambda_{d=2}=3g^2/\left(4\pi v_F\sqrt{c_sr}\right)$.
However, at higher energies $|\omega_n|>\omega_{\rm sf}$ the fermionic
spectrum exhibits non-Fermi-liquid behavior as the self-energy
(\ref{nsse}) on the imaginary axis becomes proportional to the square root of the frequency,
$\Sigma^{(p)} (i\omega_n)\propto i \, \mathrm{sign}(\omega_n)|\omega_n|^{1/2}$. 

For our subsequent analysis, it will be important to determine the
fermionic self-energy for arbitrary dimensions $d\leqslant 3$ using
the $\varepsilon$-expansion with the small parameter
\begin{equation}
\varepsilon =3-d.    \label{epsexp}
\end{equation}
Similarly to Eq.~(\ref{nsse}), we find the non-Fermi-liquid behavior
at high energies  
\begin{equation}
\Sigma^{(p)}  (i\omega_n) =\left\{ 
\begin{matrix}
-i\omega_n \lambda & , & {\rm if} \; |\omega_n| \ll \omega_{\text{sf}} \cr 
-i\omega_n \left|\frac{\bar{\Omega}}{\omega_n}\right|^{\varepsilon/2} & ,
& {\rm if} \; |\omega_n| \gg \omega_{\text{sf}}
\end{matrix}
\; , \right.   
\label{self energy d}
\end{equation}
that is characterized by the coupling constant 
\begin{equation}
\lambda =\left( 1-\frac{\varepsilon }{2}\right) 
\left( \frac{\bar{\Omega }}{\omega_{\mathrm{sf}}}\right)^{\varepsilon/2},
\label{cconst}
\end{equation}
and the energy scale 
\begin{equation}
\bar{\Omega }=\gamma ^{-1}
\left[ \frac{3g^2K_{d-1}}
{4vc_s^{1-\varepsilon/2} 
\left(1-\frac{\varepsilon}{2}\right)
\sin\frac{\pi\varepsilon}{2}}
\right]^{2/\varepsilon},
\label{Bd}
\end{equation}
where $K_d = 2^{1-d} \pi^{-d/2}/ \, \Gamma (d/2)$ contains the information about the surface of a unit sphere in $d$ dimensions.
On the real axis this yields in the non-Fermi liquid regime 
\begin{equation}
\Sigma^{(p)} (\omega) =-\omega 
\left|\frac{\bar{\Omega}}{\omega}\right|^{\varepsilon /2}
e^{i \pi \varepsilon\mathrm{sign}(\omega)/4}.
\end{equation}
For $d=3$, we find 
$\Sigma(i\omega_n)=  -i \frac{3g^2}{8 \pi^2 v_F c_s} \omega_n \log(\omega_0/|\omega_n|)$ 
with the characteristic   frequency $\omega_0 = c_s q_0^2 /\gamma$, where $|\mathbf q|< q_0$ is the bosonic momentum cutoff. On the real axis this becomes 
\begin{equation}
\Sigma^{(p)} (\omega) \propto - \omega \log \frac{\omega _{0}}{|\omega|}
-i\frac{\pi }{2} |\omega|.  
\label{d=3}
\end{equation}
Note, that Eq.(\ref{d=3}) holds only for momenta on the hot lines, in
contrast to Ref.~\onlinecite{MFL}, where within the marginal
Fermi-liquid phenomenology the same frequency dependence is assumed
everywhere on the Fermi surface. 

The above results for the normal-state fermionic dynamics demonstrate
that the upper critical dimension for non-Fermi liquid behavior of the
fermionic spectrum at the hot-spots is $d_{\text{uc}}=3$. Near three dimensions we can develop an $\varepsilon$-expansion which is controlled
for arbitrary $N$. As we show below, in the limit $\varepsilon \rightarrow 1$ (i.e. for $d=2$) the
$\varepsilon$-expansion is not reliable anymore. One might hope that
an expansion with respect to $1/N$ can be developed. As shown in
Refs.~\onlinecite{SSLee2009,Metlitski2010} for $d=2$ and gapless
fermions in the normal state, the usual loop expansion does not
correspond to an expansion in $1/N$, making a controlled expansion in
$1/N$ a complicated task, amounting to the summation of all planar
diagrams. An important question is whether the
dynamics in the superconducting state, where fermions are gapped, is
still plagued by similar problems.

\subsection{Pairing instability}

In order to investigate the emergence of the resonance mode, we will
consider the spin-fermion model deep in the superconducting state. For this we need an estimate of the superconducting gap amplitude at low temperatures. Here, we obtain this quantity by combining the numerical solution of Ref.~\onlinecite{AbanovEPL01} with the linearized gap equations near $T_c$ (for varying dimension $d$). Since these gap equations were solved
elsewhere~\cite{Bonesteel1996,Son1999,AbanovEPL01,AbanovEPL01b,Chubukov2005,Moon2010},
we merely summarize the key results to make the article self-contained and in order to introduce the notation used throughout this paper.

In the superconducting state we express anomalous averages through the
self-energy $\Phi_k$ and determine this quantity, along with the
associated gap function $\Delta_k=\Phi_k/Z_k$ self-consistently. Since
the dominant contribution to the bosonic self-energies comes from the
hot spots, one obtains for these momenta
$\Delta_\mathbf{k+Q}=\pm\Delta_\mathbf{k}$. In the case of cuprate
superconductors, the minus sign corresponds to $d$-wave pairing. In
the case of the iron-based superconductors, the minus sign corresponds
to the $s_\pm$ state or a d-wave state, depending on the typical spin-momentum vector $\mathbf Q$.

For $d=2$, the gap equation determining $\Delta_\mathbf{k}$ was solved
in Refs.~\onlinecite{AbanovEPL01,AbanovEPL01b}. It was found that the
amplitude of the gap function $\Delta(T\ll T_c)$ is proportional to
the instability temperature $T_c$, with $2\Delta /T_c\simeq 5$. Thus,
in what follows we will merely determine $T_{c}$ and use it as an
estimate for the gap amplitude in the superconducting state. Related
pairing problems with singular pairing interactions were discussed in
the context of gauge-field induced pairing in quantum-Hall double
layers~\cite{Bonesteel1996}, color superconductivity~\cite{Son1999}
and the strong coupling behavior in problems with massless boson
exchange in three dimensions~\cite{Chubukov2005}. Quantum-critical
pairing with power-law dependence of the pairing interaction were studied in
Ref.~\onlinecite{Moon2010}. In what follows we summarize the key
results for quantum-critical pairing as a function of $\varepsilon$.

The one-loop fermionic self-energy matrix in Nambu-space follows from
Eq.~(\ref{sfsc}):
\begin{equation}
\widehat\Sigma_k=g^2\int_q\sum_{i=1}^3 \hat\alpha^i\chi_q  
\widehat{\cal G}_{k-q} \hat\alpha^i 
= 3 g^2 \int_q \chi_q \widehat{\cal G}_{k-q}  ,
\end{equation}
Using Eq.~(\ref{Sigmadef}) and this self-energy we obtain the functions:
\begin{align}
Z_k &= 1-\frac{3g^2}{2 i\omega_n}\int_q\chi_q  \bigl[ {\cal G}_{k-q}^{(p)} +{\cal G}_{k-q}^{(h)} \bigr],  
\notag \\
\delta \epsilon_k &= \frac{3g^2}{2 }\int_q\chi_q  \bigl[ {\cal G}_{k-q}^{(p)} -{\cal G}_{k-q}^{(h)} \bigr] ,
\nonumber\\
\Phi_k &=3g^2\int_q\chi_q{\cal F}_{k-q}.
\end{align}%
The normal and anomalous Green's function in the superconducting state are thus given by Eq.~(\ref{propagator}). The self-energies near the hot spots are weakly momentum-dependent and therefore we assume the dispersion correction $\delta \epsilon_k=0$ for the determination of the superconducting transition temperature, because the frequency dependence is dominant in the $Z_k \approx Z(i \omega_n)$ term. Integrating over fermionic energies $\varepsilon
_{\mathbf{k}}$ then yields the linearized Eliashberg equations \cite{eli,car} [noting that $\phi_{\mathbf{k_F+Q}}(i \omega_n) =-\phi_{\mathbf{k_F}}(i \omega_n)$ ]
\begin{align}
\begin{split}
\Phi_{\mathbf{k_F}}(i \omega_n) &=
\pi T\sum_m {\cal D}(i\omega_n-i\omega_m) \frac{\Phi_{\mathbf{k_F+Q}}(i\omega_m)}
{|\omega_m| Z(i\omega_m)} , 
\\
Z_{\mathbf{k_F}}(i\omega_n) &= 1 +
\frac{\pi T}{\omega_n} \sum_m{\cal D}(i\omega_n-i\omega_m) \mathrm{sign}(\omega_m) .
\end{split}
\label{eleq}
\end{align}
that determine $T_c$. The self-energies are evaluated at the momenta
$\mathbf{k}_F$ and $\mathbf{k}_F+\mathbf{Q}$, which is suppressed in the
notation.  The
effective coupling function in Eq.~(\ref{eleq}) is given by the 
integral
\begin{equation}
{\cal D}(i\omega_n) =
\frac{3g^2}{4\pi^2v_F} \int\frac{d^{d-1}q_\|}{(2\pi)^{d-1}}
\frac{1}{r+\gamma |\omega_n| +c_sq_\|^2}.
\end{equation}
Here integration over momenta is performed over the $d-1$
components of the bosonic momentum that are parallel to the Fermi
surface\cite{Chubukov2005}. The result of the
integration is given by
\begin{equation}
{\cal D}(i\omega_n) = \frac{1-\varepsilon /2}{2\pi } 
\left[ \frac{\bar{\Omega}}{\omega_{\rm sf}+|\omega_n|} \right]^{\varepsilon/2},
\end{equation}
with the energy scale $\bar{\Omega }$ defined in Eq.~(\ref{Bd}). 

\begin{figure}
\begin{center}
\includegraphics[width=0.6\textwidth]{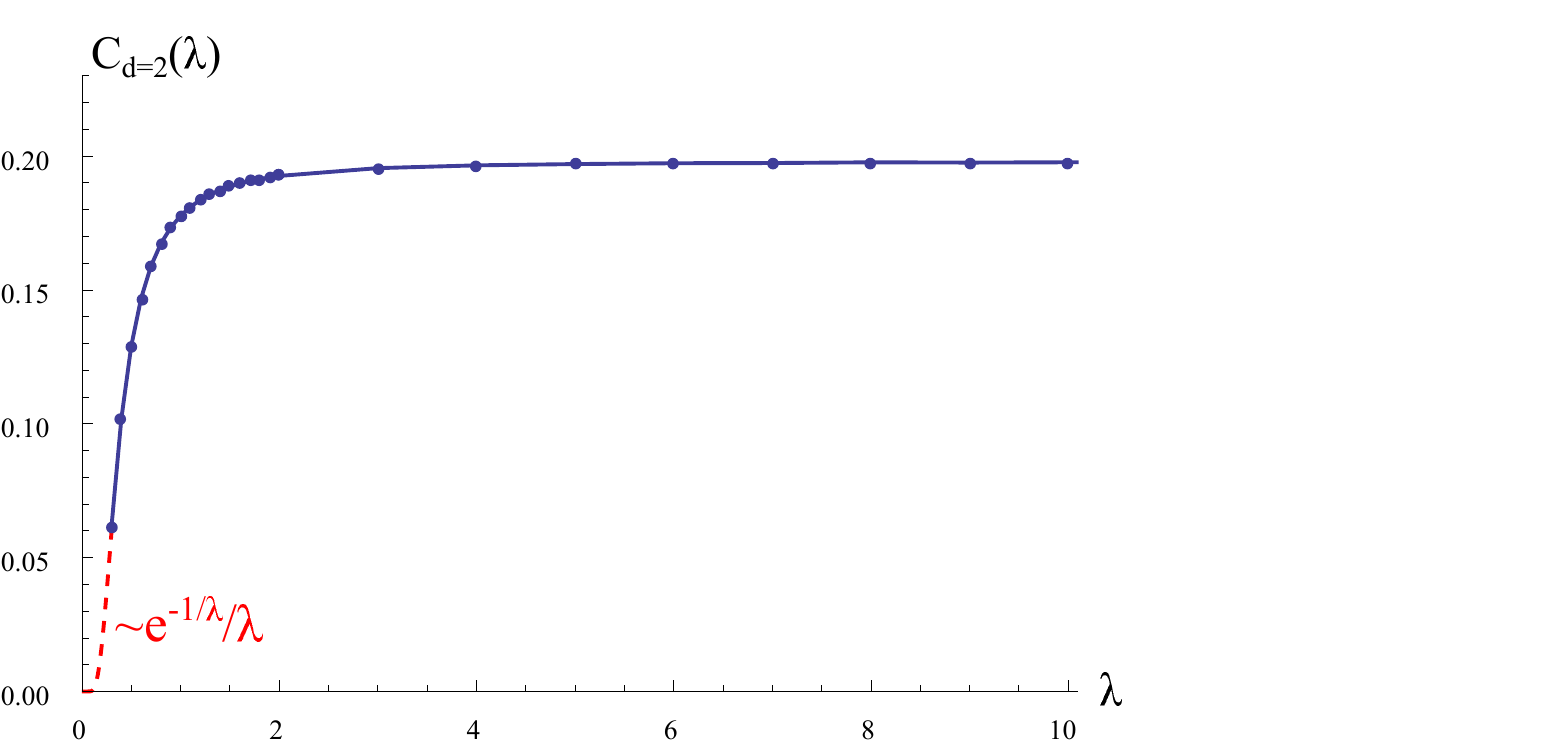}
\caption{$\lambda$ dependence of the universal function
  $C_\epsilon(\lambda)$ for $d=2$. The dashed curve is an exponential
  fit to the expected weak-coupling behavior and the dots represent
  the numerical values of the strong coupling calculation.}
\label{C2}
\end{center}
\end{figure}

The Matsubara gap function $\Delta_n=\Phi(\omega_n)/Z(\omega_n)$ obeys
the linearized equation
\begin{equation}
\Delta_n=\pi T\sum_m{\cal D}(i\omega_n-i\omega_m)
\left[ \frac{\Delta_m}{\omega_m}-\frac{\Delta_n}{\omega_n}\right]
\mathrm{sign}(\omega_m). 
\end{equation}
It is convenient to bring this equation to the form
\begin{eqnarray}
&&
\Delta_n=\frac{1-\varepsilon /2}{2\pi}
\left[ \frac{\bar{\Omega}}{2\pi T}\right]^{\varepsilon /2}
\sum_m \frac{\mathrm{sign}(2m+1)}{\left( \frac{\omega_{\mathrm{sf}}}{2\pi T}
+|2n-2m| \right)^{\varepsilon /2}}  
\notag \\
&&
\nonumber\\
&&
\quad\quad\quad
\times 
\left[ \frac{\Delta_m}{2m+1}-\frac{\Delta_n}{2n+1}\right] 
\mathrm{sign}(2m+1).
\end{eqnarray}
\begin{figure}
\begin{center}
\includegraphics[width=0.4\textwidth]{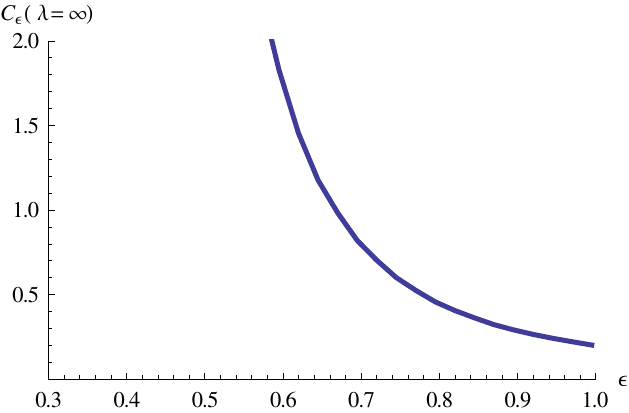}
\caption{$\varepsilon$ dependence of the universal function
  $C_\epsilon(\infty)$.}
\label{Cepsilon}
\end{center}
\end{figure}
At the quantum critical point, where $\omega _{\mathrm{sf}}=0$, it
holds that the transition temperature must be determined by a critical
value of the coefficient in front of the Matsubara sum. Then the ratio
$\bar{\Omega}/T_c$ should take a universal value yielding $T_c\simeq
\bar{\Omega}$. Away from the critical point, the transition
temperature may be written in the form
\begin{equation}
T_c=\bar{\Omega }C_\varepsilon(\lambda),    
\label{Tc}
\end{equation}
with universal function $C_\varepsilon (\lambda)$ of the dimensionless
coupling constant $\lambda$ defined in Eq.~(\ref{cconst}). For
$\lambda\ll 1$, we recover the BCS behavior
$T_c\propto\bar\Omega\lambda^{-1}\exp(-1/\lambda)$.  However, in this
regime magnetic correlations are so short-ranged that our continuum
theory is no longer the appropriate starting point. On the other hand,
if the coupling constant is larger than unity, the pairing is
quantum-critical and $T_{c}\simeq\bar\Omega$. In Fig.~\ref{Cepsilon} we show the numerical dependence of the strong coupling limit $C_\varepsilon(\infty)$ as a function of the dimensional expansion parameter $\varepsilon$. From the numerical
solution of the gap equation we find for the case of two dimensions $C_{d=2}(\infty)=0.198$(the
full numerical dependence on $\lambda$ is shown in
Fig.~\ref{C2}). Although these results are obtained in the limit of
large $\lambda$, the calculation is well controlled in the limit of
small $\varepsilon$. In our subsequent analysis we therefore use
$\Delta \simeq \bar{\Omega }$ in the regime of strong magnetic
correlations (i.e. for $\lambda >1$).

Finally, for $d=3$ the power-law dependence of the transition
temperature~(\ref{Tc})  becomes
\begin{equation}
T_{c}\left( d=3\right) \propto \exp \left( -\pi /g\right), 
\qquad (\lambda \gg 1),
\end{equation}
which is fully consistent with earlier
results~\cite{Son1999,Chubukov2005}. In a recent publication~\cite{Chubukov2013}, it was shown that  momentum dependent self-energies   correct the numerical values of Eq.~(\ref{Tc}), yet do not modify  the $\bar  \Omega$ dependence. Here, we ignore these effects in the determination of the pairing amplitude. This is justified since we are only interested in order  of magnitude	of the pairing gap.

\section{Analysis of the resonance mode}
Now we discuss the implications of the above picture for the behavior
of the resonance mode in the vicinity of a magnetic quantum-critical
point.

The analysis of the resonance mode as a spin-exciton in the superconducting
state, caused by scattering between quasiparticles and the condensate, was
investigated in Refs.~\onlinecite%
{Abanov2000,Eschrig2000,Abanov2001,Eschrig2002,Abanov2002,Eschrigreview} and based on the determination of the leading contribution to
the bosonic spin self-energy.
Key concepts for the emergence of the resonance mode can  be carried over from the
analysis of the leading order terms. To this end, we follow Abanov and
Chubukov~\cite{Abanov2000} and discuss the emergence of a resonance mode in
the superconducting state.  Generally, the imaginary part of bosonic self
energy $\Pi _{\mathbf{Q}}\left( \omega \right) $ vanishes at $T=0$ for
frequencies $\left\vert \omega \right\vert <2\Delta $, where $\Delta = |\Delta_{\mathbf{k_F}}|$ is the amplitude of the superconducting gap at the hot spot. \ Within weak coupling theory holds that  ${\rm Im} \Pi _{\mathbf{Q}}\left( \omega \right) 
$ grows continuously at $\omega =\pm 2\Delta $ according to $\gamma \sqrt{%
\left( \left\vert \omega \right\vert -2\Delta \right) \Delta }$ if $\Delta _{
\mathbf{k_F}}$ and $\Delta _{\mathbf{k_F}+\mathbf{Q}}$ have the same
phase. However, as soon as the phases of $\Delta _{\mathbf{k_F}}$ and $\Delta _{\mathbf{k_F}+\mathbf{Q}}$ differ, ${\rm Im} \Pi _{\mathbf{Q}}\left( \omega \right) $ becomes discontinuous at $\omega=2 \Delta$. A key quantity for our analysis is therefore the
height of this discontinuity:%
\begin{equation}
D\equiv \lim_{\delta \rightarrow 0^{+}}{\rm Im} \Pi _{\mathbf{Q}}(2\Delta
+\delta ).  \label{disc}
\end{equation}%
Once $D>0$ the discontinuity in the imaginary part of $\Pi _{\mathbf{Q}%
}\left( \omega \right) $ translates into a logarithmic divergence of its
real part at $2\Delta $: 
\begin{equation}
\text{Re}\Pi _{\mathbf{Q}}(\omega \simeq 2\Delta )=-\frac{D}{\pi }\ln \left( 
\frac{\left\vert \omega -2\Delta \right\vert }{2\Delta }\right) .   \label{RePioneloop}
\end{equation}
\begin{figure}
\centering
\includegraphics[width=0.48\textwidth]{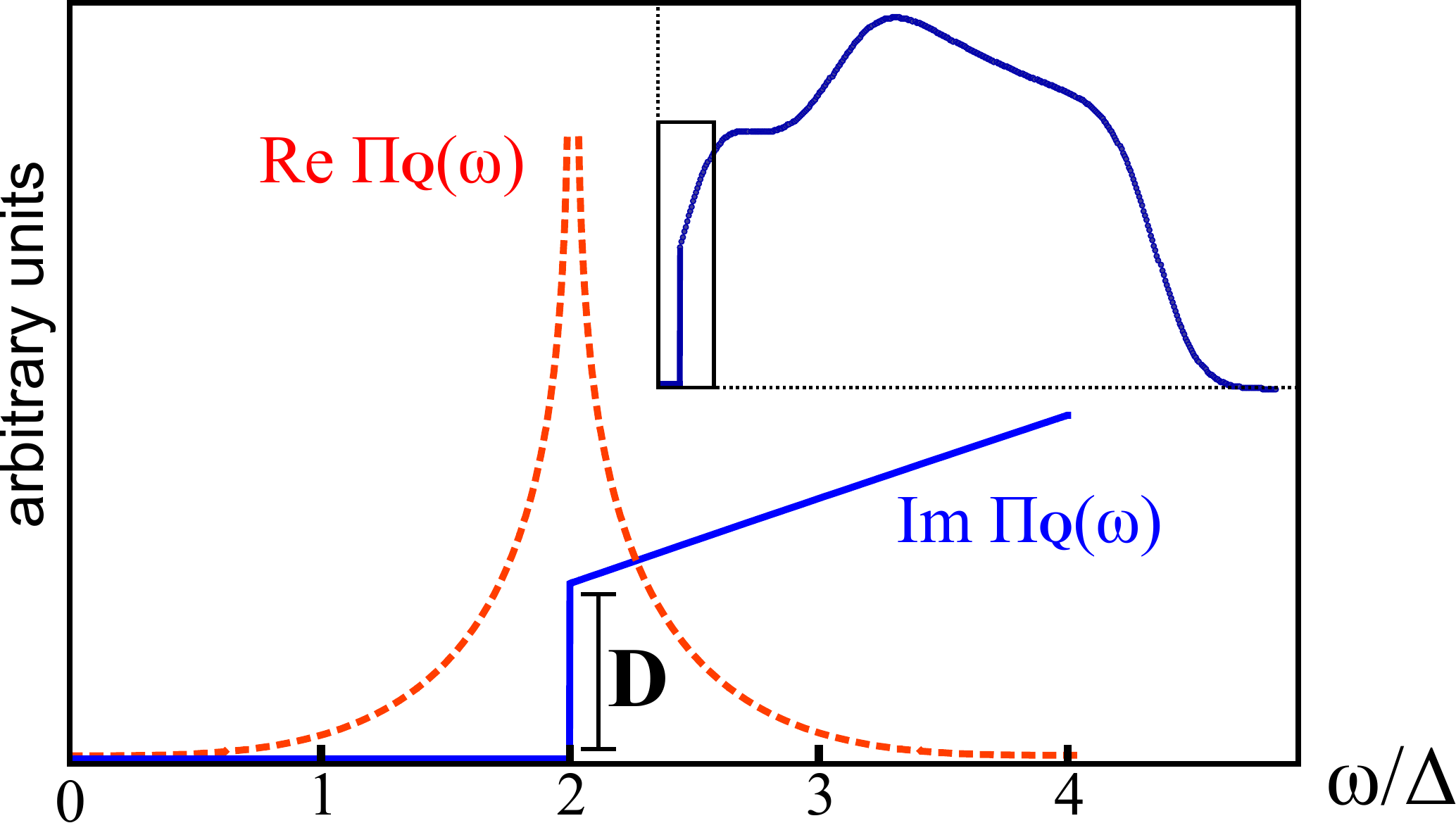}
\caption{Schematic plot of the real and imaginary part for the bosonic self-energy with discontinuity $D > 0$ at $\omega = 2 \Delta$ for a superconductor with varying phase $\Delta_{\mathbf{k_F}+\mathbf{Q}} = - \Delta_{\mathbf{k_F}}$. On the upper right side we plot a numerical one-loop analysis of the imaginary part  for a realistic spectrum of Bi-2212, where we used experimental parameters of Ref.~\cite{Kordyuk2003}.}
\label{PI2}
\end{figure}

\noindent Within one-loop approximation the susceptibility~(\ref{spinsus}) with self-energy~(\ref{RePioneloop}) yields Eq.~(\ref{r1}). The resulting energy of the resonance mode is 
\begin{equation}
\Omega _{\text{res}}=2\Delta \left( 1-e^{-\frac{\pi r}{D}}\right) 
\label{r2}
\end{equation}%
 with spectral weight%
\begin{equation}
Z_{\text{res}}=\frac{2\pi ^{2}\Delta }{D}e^{-\frac{\pi r}{D}}=\frac{2\pi
^{2}\Delta }{D}\left( 1-\frac{\Omega _{\text{res}}}{2\Delta }\right) .
\label{r3}
\end{equation}%
The resonance energy is bound to occur below the particle-hole continuum
that sets in at $\omega =2\Delta $, while the imaginary part of the incoherent
contributions $\chi _{\mathbf{Q}}^{\mathrm{inc}}\left( \omega \right) $ vanishes for $%
\left\vert \omega \right\vert <2\Delta $, see Fig.~\ref{PI2}. Below we will see that at one-loop
order, the discontinuity is given by $D_{0}=\pi \gamma \Delta $ such that $%
Z_{\chi }^{0}=\frac{2\pi }{\gamma }\left( 1-\frac{\Omega _{\text{res}}}{%
2\Delta }\right) $.  The above results are correct as long as $\Omega_{\text{res}}$ is of order $\Delta$. However, in the limit $\lambda \rightarrow \infty$ it was shown that~\cite{Abanov2000} $\Omega_{\text{res}} \simeq \sqrt{\omega_{\text{sf}} \Delta} \simeq \Delta/\lambda$ is determined by the leading low-frequency dependence of $\text{Re} \Pi_{\mathbf{Q}}(\omega) \simeq \gamma \omega^2/\Delta$. Here, we focus on the former regime.

Our analysis of corrections to the spin-susceptibility that go beyond the
leading order still yields that ${\rm Im} \Pi _{\mathbf{Q}}\left( \left\vert
\omega \right\vert <2\Delta \right) =0$ .  The emerging discontinuity $D$ is then solely responsible for all of the qualitative features of the model, including Eqs.~(\ref{r1}) and~(\ref{r2}). In order to determine the self-energy $%
\Pi _{\mathbf{q}}\left( \omega \right) $ of the collective spin excitations,
we start from the action Eq.~(\ref{sfsc}) and integrate out the gapped
fermions, leading to a theory of the collective spin modes: 
\begin{align}
S& =\frac{1}{2}\int_{q}\chi _{q,0}^{-1}\mathbf{S}_{q}\cdot \mathbf{S}_{-q}-%
\frac{1}{2}\text{tr}\ln \left( -\beta \hat{G}_0^{-1}\right)  \notag \\
& \quad +\frac{1}{2}\sum_{n=1}^{\infty }\frac{g^{n}}{n}\text{tr}\left[ ( \hat {G}_0 \hat{\bm \alpha} \cdot 
\mathbf{S})^{n}\right] .
\end{align}

\noindent The usual skeleton expansion follows from
expanding the logarithm. The overall factor $\frac{1}{2}$ in front of the
second term is a consequence of the fact that $\Psi $ and $\Psi ^{\dagger }$
are not independent Grassmann fields since we had to extend the Nambu spinor due to the spin-changing interaction and one has to be careful in integrating out the fermionic degrees of freedom. Here, we use the identity~\cite{Greiner}
\begin{align}
\int D \bm \eta \,  e^{- \frac{1}{2} \bm \eta^T \hat A  \bm \eta } = \sqrt{\text{det}(\hat A)} \, ,  \label{aio1}   
\end{align}
where $\bm \eta$ is a Grassmann vector and $\hat A$ a quadratic matrix. It is possible to write our path integrals in this form by using the symmetry
\begin{align}
\bar \Psi_{k'} = \Psi_{-k'}^T  \hat O  \qquad \qquad   \text{with } \hat O = \biggl(\begin{matrix} 0 & \mathds 1_2 \\ \mathds 1_2 & 0 \end{matrix} \biggr)  .  \label{aio2} 
\end{align}
We then obtain
\begin{align}
\int D [\Psi_k] & \, ,e^{- \frac{1}{2} \sum_{k,k'} \bar \Psi_{k'} \hat A_{k',k}  \Psi_k}   \nonumber \\
&= \int D[\Psi_k]e^{- \frac{1}{2} \sum_{k,k'}  \Psi_{k'} (\hat O \hat A_{-k',k})  \Psi_k}   \nonumber \\
&= \int D[\Psi_k]e^{- \frac{1}{2} \sum_{k,k'}  \Psi_{k'} (\hat O \hat A'_{k',k})  \Psi_k}  \nonumber \\
&= \sqrt{\text{det}(\hat O \hat A')} =  e^{\frac{1}{2} \text{tr} \ln(\hat A')}  \, ,  \label{aio3}
\end{align}
where we define $A_{k',k}'=A_{-k',k}$ and use that the determinant of $\hat O$ is 1.  The expansion of the logarithm leads to the known perturbation series and it is easy to see that we are  allowed to replace $\hat A'$ with the initial matrix $\hat A$. In summary, the only difference to the usual integration over two independent Grassmann fields is the factor $\frac 1 2$ in front of the $\text{tr} \ln (\ldots)$ term of the effective action. 

In the superconducting state the propagator matrix $\hat G_0$ should be replaced by $\hat{\mathcal{G}}_k$ to make the theory self-consistent. 

\subsection{Resonance mode at one-loop}

Within the one-loop approximation the bosonic self-energy  will be of order $g^{2}$
. The corresponding contribution to the action  is
given by 
\begin{align}
\delta S{}^{(2)}& =\frac{g^{2}}{4}\text{tr}\left[ ( \hat {\cal G} \hat{\bm \alpha} \cdot 
\mathbf{S})^{2}\right]  \notag \\
& =\frac{g^{2}}{4}\int_{k,q}S_{-q}^{i}S_{q}^{j}\text{tr}_{\sigma }(\hat{\cal G}%
_{k}\alpha ^{i}\hat{\cal G}_{k+q}\alpha ^{j})  \notag \\
& =-\frac{1}{2}\int_q \mathbf{S}_{-q}\cdot \mathbf{S}_{q}\,\Pi _{\mathbf{q}%
}^{\left( 2\right) }(i\omega _{n}).
\end{align}

\noindent Here, the one-loop boson self-energy  is
\begin{align}
\Pi _{\mathbf{q}}^{\left( 2\right) }(i\omega _{n})&=-2g^{2}\int_{k}\left(
{\cal G}_{k}^{(p)}{\cal G}_{k+q}^{(p)}+{\cal F}_{k} {\cal F}_{k+q}^{\ast }\right)  \nonumber  \\
&= \begin{matrix} \vspace{-2mm}
\includegraphics[width=2cm]{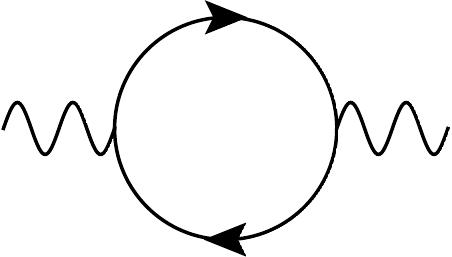}
\end{matrix}+\begin{matrix} \vspace{-2mm}
\includegraphics[width=2cm]{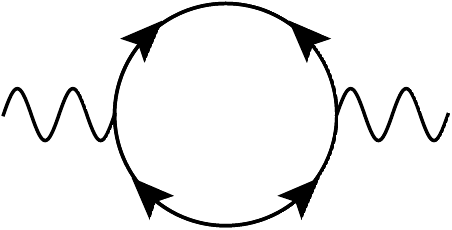} 
\end{matrix} \, .   \label{pol1}
\end{align}
The self-energy $\Pi _{\mathbf{q}}^{\left( 2\right) }(i\omega _{n})$ is $g^{2}$ times the
spin-susceptibility of fermions in the BCS theory. Using the standard mean-field approach (which here amounts to setting $Z_k=1, \delta \epsilon_k=0$ and  $\Phi_k= \Delta_{\mathbf k} $  constant in frequency) we find that on the real axis
\begin{eqnarray}
{\rm Im} \Pi _{\mathbf{q}}^{\left( 2\right) }(\omega ) &=&2g^{2}\int \frac{%
d^{d}k}{\left( 2\pi \right) ^{d}}\int \frac{d\varepsilon}\pi \left( f\left(
\varepsilon \right) -f\left( \varepsilon +\omega \right) \right)  \notag \\
&&\times \biggl[ {\rm Im} {\cal G}_{\mathbf{k}}^{(p)}\left( \varepsilon \right) {\rm Im} %
{\cal G}_{\mathbf{k+q}}^{(p)}\left( \varepsilon +\omega \right)   \notag \\
&& \hspace{3mm}+ {\rm Im} {\cal F}_{\mathbf{k}}\left( \varepsilon \right) {\rm Im} {\cal F}_{%
\mathbf{k+q}}^{\ast }\left( \varepsilon +\omega \right) \biggr] .  \label{impi12}
\end{eqnarray}%
The fermionic propagators in the superconducting state can be written as
\begin{equation*}
{\cal G}_{\mathbf{k}}^{(p)}(\omega )=\frac{u_{\mathbf{k}}^{2}}{\omega
+i0-\xi _{\mathbf{k}}}+\frac{v_{\mathbf{k}}^{2}}{\omega
+i0+\xi _{\mathbf{k}}}
\end{equation*}
with the coherence factors $u_{\mathbf k}^2 =1/2(1+\epsilon_{\mathbf k}/\xi_{\mathbf k}),v_{\mathbf k}^2 =1/2(1-\epsilon_{\mathbf k}/\xi_{\mathbf k})$ and superconducting dispersion $\xi_{\mathbf{k}} = \sqrt{\epsilon_{\mathbf k}^2 + |\Delta_{\mathbf k}|^2} \approx \sqrt{\epsilon_{\mathbf k}^2+ \Delta^2}$ at the hot spots.  For zero temperature and positive $\omega >0$ Eq.~(\ref{impi12}) yields 
\begin{align}
{\rm Im} \Pi _{\mathbf{q}}^{\left( 2\right) }(\omega )& =2\pi g^{2}\int 
\frac{d^{d}k}{\left( 2\pi \right) ^{d}}   \biggl[ u_{\mathbf{k}
}^{2}v_{\mathbf{k+q}}^{2}  \label{smpaper12} \\
& \qquad  - u_{\mathbf k}v_{\mathbf k}u_{\mathbf {k+q}}v_{\mathbf {k+q}} \biggr]
\delta (\omega -\xi _{\mathbf{k}}-\xi _{\mathbf{k+q}}).  \notag
\end{align}
Since the self-energy for negative frequencies $\omega <0$ can be easily obtained from $%
{\rm Im} \Pi _{\mathbf{q}}(-\omega )=-{\rm Im} \Pi _{\mathbf{q}}(\omega )$ we will restrict further calculations to $\omega>0$.
To analyze the resonance mode near the antiferromagnetic ordering vector, we evaluate 
$\Pi _{\mathbf{q}}^{\left( 0\right) }(\omega )$ at $%
\mathbf{q}=\mathbf{Q}$. The integral in Eq.~(\ref{smpaper12}) is dominated by 
fermions near the hot spots on the Fermi surface.  Consequently, ${\rm Im} \Pi _{\mathbf{Q}%
}^{\left( 0\right) }(\left\vert \omega \right\vert <2\Delta )=0$ leading to a spin gap in the spectrum of the resonance mode. 

Near the $2\Delta$~threshold the imaginary part of the bosonic self-energy~(\ref{smpaper12}) exhibits a discontinuity. within the one-loop calculation, the height of the discontinuity is given by~\cite{Abanov2000}%
\begin{equation}
D_{0}=\pi \gamma \Delta ,  \label{Do}
\end{equation}%
with $\gamma $ from Eq.~(\ref{gamma}). This result occurs for sign-changing gap  $\Delta _{\mathbf{%
k}_{F}}=-\Delta _{\mathbf{k}_{F}+\mathbf{Q}}$.  It is straightforward to analyze Eq.~(\ref
{smpaper12}) \ for the more general pairing-state with $\Delta _{\mathbf{k}%
_{F}}=\Delta _{1}e^{i\varphi _{\mathbf{1}}}$ and $\Delta _{\mathbf{k}_{F}+%
\mathbf{Q}}=\Delta _{2}e^{i\varphi _{2}}$. Now the discontinuity occurs at $%
\omega =\Delta _{1}+\Delta _{2}$ and is given by 
\begin{equation}
D_{0}=\pi \gamma \sqrt{\Delta _{1}\Delta _{2}}\sin ^{2}\left( \frac{\varphi
_{1}-\varphi _{2}}{2}\right) .    \label{D0}
\end{equation}%
The resonance occurs as long as the gap amplitude of both states connected by $\mathbf{Q}$ is
finite and the phases of the pairing states are distinct. For $\omega
> 2\Delta $ the imaginary part of the boson self-energy will grow linearly with $\omega $ until it saturates
when it reaches the band-width of the fermions. This general behavior can be seen in the numerical plot shown in Fig.~\ref{PI2}.

\subsection{Higher order corrections to the resonance mode}
In Ref.~\onlinecite{Abanov2003} it was shown that for $d=2$, vertex corrections  in the spin-fermion model lead to logarithmic divergences in the normal state. Evaluating the spin-fermion vertex corrections in the superconducting state one finds that this logarithmic divergency is cut off at the scale of the superconducting gap $\Delta $. On the other hand, an analysis of the gap equation for spin-fluctuation-induced pairing yields for arbitrary $d<3$ that quantum critical excitations with $\omega>\omega_{\text{sf}}$ are important for the value of the transition temperature $T_{c}$. \ Therefore, we examine the higher orders in perturbation theory in more detail. Specifically, we are interested in corrections  to the discontinuity $D$ of Eq.~(\ref{disc}). Diagrammatically these corrections are given by 
\begin{equation}
\delta \Pi_{\mathbf Q} (i\omega _{n})= \begin{matrix} \includegraphics[width=0.14\textwidth]{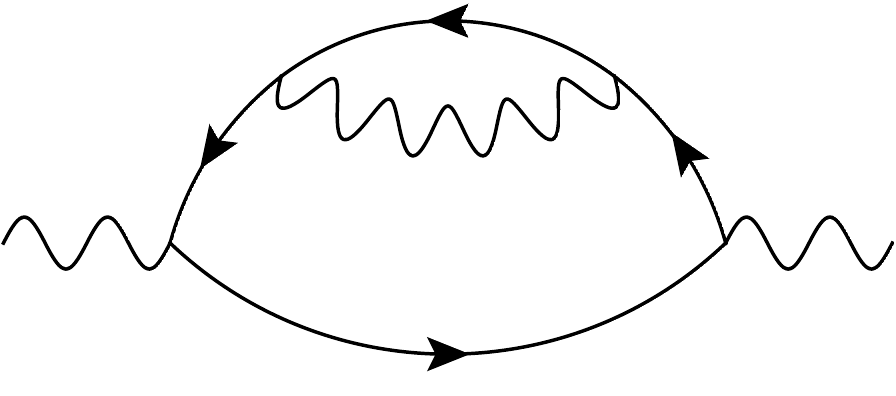} \end{matrix}+ \begin{matrix} \includegraphics[width=0.12\textwidth]{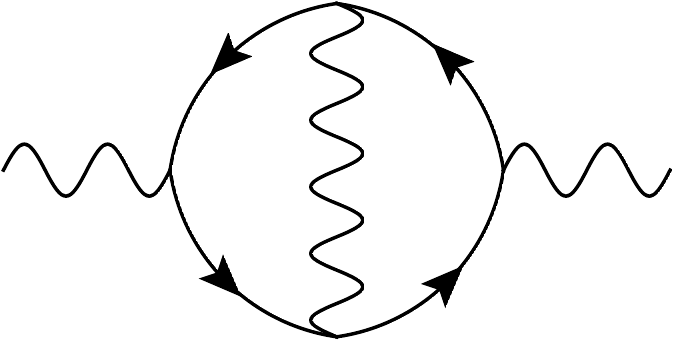}   \end{matrix}  \label{Feynmancorr}
\end{equation}
Here, the wavy lines correspond to Eq.~(\ref{spinsus}) with the one-loop bosonic self-energy. The  fermionic lines are the mean-field Green's functions used in the previous section.

\subsubsection{Self-energy corrections}
The first diagram in Eq.~(\ref{Feynmancorr}) takes into account the self-energy corrections to the fermionic Green's functions. To the leading order these are calculated in Appendix~\ref{selfenergiesinsc}. The imaginary parts of the normal and anomalous self-energies
\begin{align}
\begin{split}
\Sigma_{\textbf{k}}^{(p)}(\omega) &= \includegraphics[width=1.7cm]{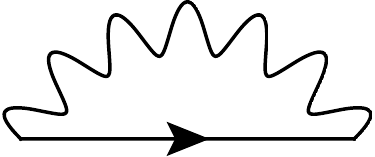}  \\
\Phi_{\textbf{k}}(\omega) &= \includegraphics[width=1.7cm]{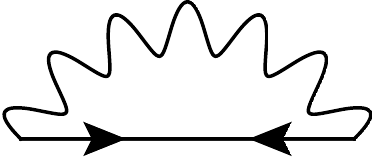}  
\end{split}    \label{secfeyn}
\end{align}
 are gapped near the hot spots for frequencies $| \omega | < \Delta+\Omega_{\text{res}}$, see also Ref.~\onlinecite{Abanov2001}. Excluding the strong coupling limit $\lambda \gg 1$ [see the discussion following Eq.~(\ref{r3})] we find that the excitations around $\omega \sim \pm \Delta$ are well separated from the continuum yielding sharp quasiparticle resonances. The minimal excitation energy $\Delta'$ is determined by the real part of the self-energy~(\ref{secfeyn}) and may be smaller than the mean-field gap $\Delta$ at the hot spots. For zero temperature we can evaluate~(\ref{impi12}) for external momentum $\mathbf{Q}$ to
 \begin{align}
\text{Im} \Pi_{\mathbf{Q}}^{(2)}(\omega) &= 2 g^2 \int \frac{d^d k}{(2\pi)^d} \int_{\Delta'}^{\omega-\Delta'} \frac{d\nu}{\pi}    \nonumber \\
&\times \biggl[ {\rm Im} {\cal G}_{\mathbf{k}}^{(h)}\left( \nu \right) {\rm Im} %
{\cal G}_{\mathbf{k+Q}}^{(h)}\left( \nu+\omega \right)   \notag \\
& \hspace{3mm}+ {\rm Im} {\cal F}_{\mathbf{k}}\left( \nu \right) {\rm Im} {\cal F}_{%
\mathbf{k+Q}}^{* }\left( \nu +\omega \right) \biggr] .   \label{pi15}
 \end{align}
Obviously, there is still a spin gap of $2 \Delta'$, which is as usual determined by twice the minimal excitation energy of the fermionic spectrum. The one-loop fermionic self-energies near the hot spot $\mathbf k_F$ are functions that depend on the dispersion on the opposite side $\epsilon_{\mathbf k+\mathbf Q}$, see Appendix~\ref{selfenergiesinsc}. Note: Our approach takes into account leading momentum and frequency corrections which arise due to the interaction of the superconducting fermions with the collective boson mode, but not two-loop corrections in the fermionic self-energy. In the considered parameter regime these momentum and frequency dependencies are weak and in~(\ref{pi15}) we see that the contributions to the discontinuity come from fermions with $\nu \approx \Delta, \nu-\omega \approx - \Delta$ which lie around the hot spot.  Therefore we expand  to leading order in momentum and frequency
\begin{align}
\begin{split}
Z_{\mathbf k}(\omega ) &=   Z_0 + Z_f (|\omega|-\Delta) + Z_m \epsilon_{\mathbf k+\mathbf Q}^2  , \\
  \Delta_{\mathbf k}(\omega ) &= \Delta + \Delta_f (|\omega|-\Delta) + \Delta_m \epsilon_{\mathbf k+\mathbf Q}^2   ,   \\
\delta \epsilon_{\mathbf k}  & = \nu_m \epsilon_{\mathbf k+\mathbf Q} ,
 \end{split}   \label{coefexp}
 \end{align}
where the coefficients $Z_0, Z_f,$ etc. are computed numerically. 
 \begin{figure}
\includegraphics[width=0.42\textwidth]{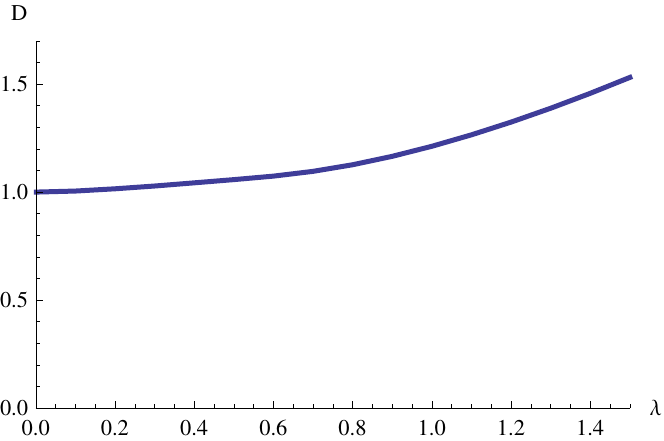}
\caption{Discontinuity $D$ containing self-energy corrections relative to one-loop jump $D_0$ for parameter range $\lambda \lesssim 1$.}
\label{selfenergycorrection}
\end{figure} 

With the help of the self-energy~(\ref{coefexp}) we find for the fermionic Green's functions in Eq.~(\ref{pi15})
\begin{align}
\text{Im}{\cal G}_{\mathbf{k}}^{(h)}(\omega) &= - \pi \biggl[\frac{v_{k}^2}{Z_k} \delta \bigl( \omega- \sqrt{\bigl( \frac{\epsilon_{\mathbf{k}}+\delta \epsilon_{\mathbf k} }{Z_k}  \bigr)^2+ |\Delta_k|^2}  \,  \bigr)      \nonumber \\
& \qquad + \frac{u_{k}^2}{Z_k} \delta \bigl( \omega+ \sqrt{\bigl( \frac{\epsilon_{\mathbf{k}}+\delta \epsilon_{\mathbf k} }{Z_k}  \bigr)^2+ |\Delta_k|^2  } \,  \bigr)  \biggr]    \nonumber
\end{align}
and correspondingly for the anomalous propagators. Here, one has to rescale the energy $\epsilon_{\mathbf{k}} \rightarrow (\epsilon_{\mathbf{k}}+\delta \epsilon_{\mathbf{k}})/Z_k$ in the coherence factors as well. Since we are interested in the discontinuity of~(\ref{pi15}) at $\omega \approx 2 \Delta$ we can expand the arguments of the two delta functions around $\nu \approx \Delta$ and evaluate the frequency integration. The  integration is greatly simplified by the usual spectrum linearization around hot spots.  Performing this analysis, we find that the minimal excitation energy of the particle-hole spectrum is still $2 \Delta$ such that the spin gap of $2 \Delta$ is unaffected by the self-energy corrections. As a result we find for  the discontinuity 
\begin{align}
D = \begin{cases} \frac{D_0}{(1-\nu_m^2)(1-\Delta_f)}  & \quad \text{for } \Delta_{\mathbf k_f+\mathbf Q} = - \Delta_{\mathbf k_F}  \\  0& \quad \text{for } \Delta_{\mathbf k_F+\mathbf Q} =  \Delta_{\mathbf k_F}   \end{cases}.   \label{Dse}
\end{align}
The ratio $D/D_0$ is shown in Fig.~\ref{selfenergycorrection} as a function of $\lambda$ for $N=8$, see Appendix~\ref{selfenergiesinsc} for further details. Since $\Delta_f,\nu_m \sim g^2$ only for $\lambda \ll 1$ the self-energy corrections are of order one for the physical regime $\lambda$ of order unity. In the limit of large $N$ the parameters $\nu_m, \Delta_f \sim 1/N$ (for arbitrary $g$), see Appendix~\ref{selfenergiesinsc}. The result suggests that using an $1/N$ expansion, the self-energy corrections can be  calculated controllably. However, previous results\cite{SSLee2009} show that for $d=2$ there are problems with the $1/N$ expansion  in the normal state with a gapless fermionic spectrum. It is unclear whether these problems persist in the superconducting state discussed here.

\subsubsection{Vertex-Corrections}
Now we turn in the examination of vertex corrections. Performing the perturbation theory in the extended spinor-space, indicated by the double-lined propagator matrices, we can express them as
\begin{widetext}
\begin{align}
\delta \Pi_{\mathbf Q}(i \omega_n) &= \begin{matrix}\includegraphics[height=2cm]{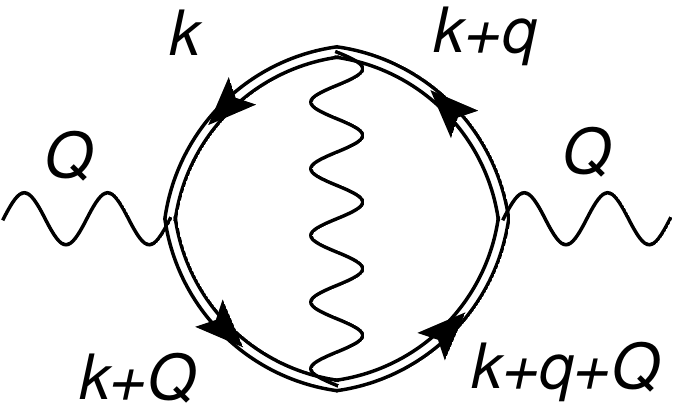} \end{matrix} = - \frac{g^4}{2}  \int_{k,q}   \chi_q \, \text{tr}( \hat{\mathcal G}_k \hat \alpha^z \hat{\mathcal G}_{k+Q} \hat{\bm \alpha}  \hat{\mathcal G}_{k+q+Q} \hat \alpha^z \hat{\mathcal G}_{k+q} \hat{\bm \alpha}) \label{smpapervc3} \\
&= -  g^4  \sum_{\{A,B,C,D\} } \int_{{k=\mathbf k, i \Omega_m} \atop {q=\mathbf q, i \nu_k}}  \chi_{\mathbf q} (i \nu_k) A_{\mathbf k}( i\Omega_m) B_{\mathbf k+ \mathbf Q}(i\Omega_m+i\omega_n)   C_{\mathbf k+\mathbf Q+\mathbf q}(i \Omega_m+i \omega_n+i \nu_k) D_{\mathbf k+ \mathbf q } (i \Omega_m+i \nu_k)   \, ,    \nonumber
\end{align}
\end{widetext}
where in the following fermionic Matsubara frequencies will be written with capital letters and bosonic ones with small letters. Due to spin rotation symmetry we can restrict ourselves to the zz-component of the bosonic self-energy.  The sum $\{A,B,C,D\}$ has to be executed over all possible combinations of Gor'kov-Nambu Green's functions with arrow conservation at each vertex
\begin{figure}
\centering
\includegraphics[width=0.47\textwidth]{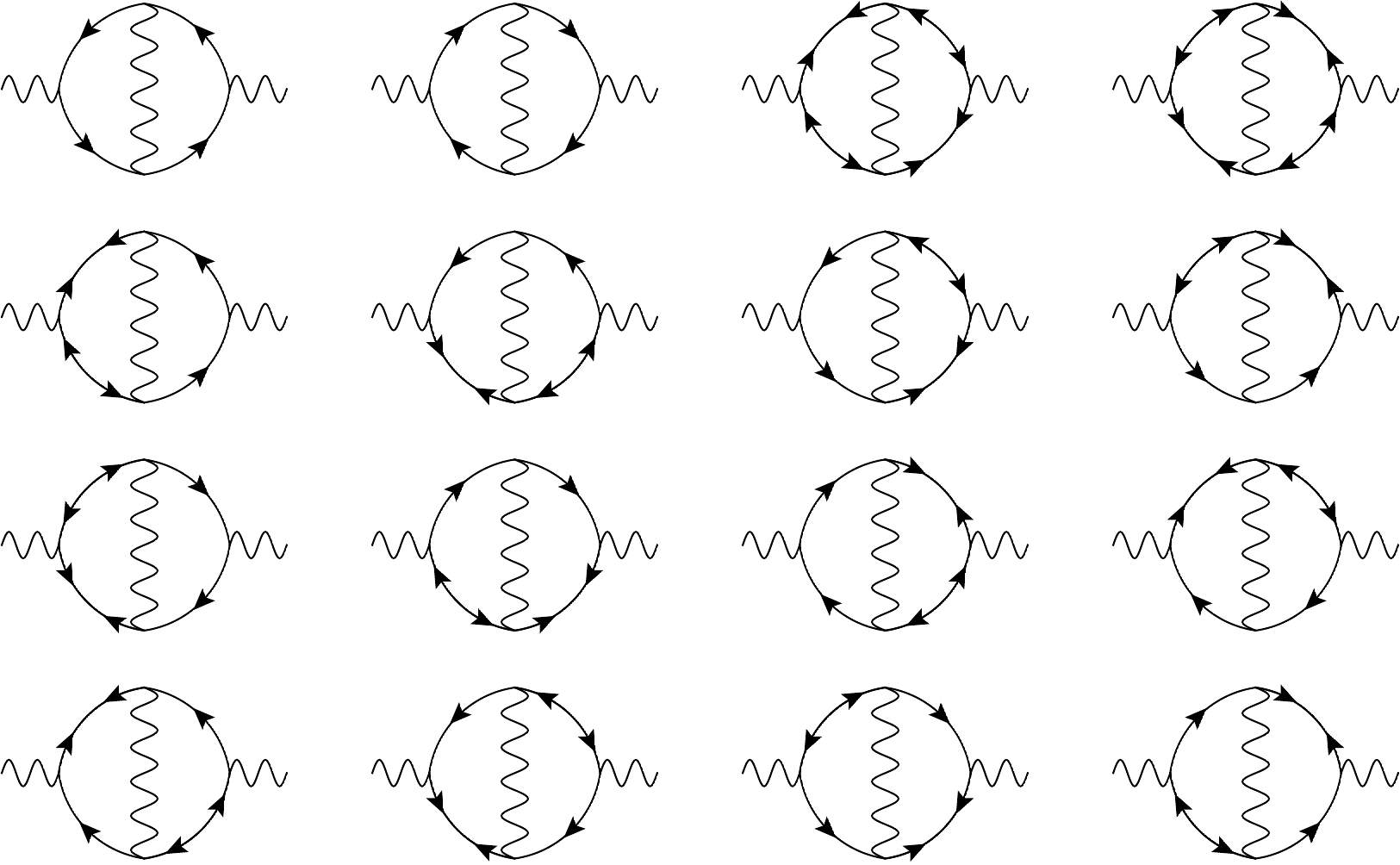}
\caption{Possible diagrams for the next-order self-energy $\delta \Pi$ in the superconducting state. Note the arrow conservation at each vertex indicating the energy-momentum conservation of the theory.}
\label{sumGFbild}
\end{figure}
\begin{align}
\begin{split}
\{A,B,C,D\} =& \mathcal G ^{(p)} \mathcal G ^{(p)} \mathcal G ^{(p)} \mathcal G ^{(p)} + \mathcal G ^{(h)} \mathcal G ^{(h)} \mathcal G ^{(h)} \mathcal G ^{(h)}   \\
&+ \mathcal F \mathcal F^* \mathcal F \mathcal F* + \mathcal F^* \mathcal F \mathcal F^* \mathcal F       \\
&+ \mathcal F \mathcal F^* \mathcal G ^{(p)} \mathcal G ^{(p)} + \mathcal G ^{(p)} \mathcal F \mathcal F^* \mathcal G ^{(p)}  \\
& + \mathcal G ^{(p)} \mathcal G ^{(p)} \mathcal F \mathcal F^*  +\mathcal F^* \mathcal G ^{(p)} \mathcal G ^{(p)} \mathcal F  \\
&+ \mathcal F^* \mathcal F \mathcal G ^{(h)} \mathcal G ^{(h)} + \mathcal G ^{(h)} \mathcal F^* \mathcal F \mathcal G ^{(h)}  \\
& + \mathcal G ^{(h)} \mathcal G ^{(h)} \mathcal F^* \mathcal F  + \mathcal F \mathcal G ^{(h)} \mathcal G ^{(h)} \mathcal F^* \\
&+ \mathcal F \mathcal G ^{(h)} \mathcal F^* \mathcal G ^{(p)} +\mathcal G ^{(p)}  \mathcal F \mathcal G ^{(h)} \mathcal F^*   \\
& + \mathcal F^* \mathcal G ^{(p)} \mathcal F \mathcal G ^{(h)}   + \mathcal G ^{(h)} \mathcal F^* \mathcal G ^{(p)}  \mathcal F  \, ,
\end{split}   \label{sumGF}
\end{align}
which are shown in Fig.~\ref{sumGFbild}. The correction to the bosonic self-energy $\delta \Pi_{\mathbf Q}^R(\omega)$ is evaluated in Appendices~\ref{appendix2} and \ref{appendix3} at $T=0$. Here, we are focusing on the correction to the discontinuity~(\ref{Do})
\begin{align}
  \delta D = \lim_{\delta \rightarrow 0^+} \text{Im} \delta \Pi_{\mathbf Q}( 2 \Delta + \delta).
  \end{align}
  
Since the lowest possible particle-hole excitations $\epsilon_{\mathbf{k}}+\epsilon_{\mathbf{k+Q}}$ connected by the magnetic ordering vector still lie directly at the hot spots and therefore the fermionic quasiparticles remain gapped with $\Delta$, we find that the spin gap in $\text{Im} \Pi_{\mathbf{Q}}(\omega)$ remains $2 \Delta$ even after taking into account  vertex corrections. 
  
  Below we  discuss the strong coupling behavior of the spin resonance in $d=2$ dimensions and a systematic approach to regularize the theory with an $\epsilon$-expansion around the upper critical dimension $d_{\text{uc}}=3$.

\subsubsection*{Strong coupling behavior for $d=2$}

The emergence of the discontinuity $D=D_{0}+\delta D$ in the imaginary part of the bosonic self-energy still hinges on the symmetry of the superconducting order parameter. In particular, we find similarly to~(\ref{Dse}) that there is no discontinuity for conventional gap symmetries $\delta D_{s-\text{wave}} =0$, see Appendix~\ref{appendix3} for details. The physical reason behind this result is the fact that due to phase space restrictions (for the  bosonic propagator being sharply-peaked around $\mathbf q \sim \mathbf{Q}$), the fermionic quasiparticles at hot spots do not couple to any gapless excitations. This argument can be extended to higher orders in perturbation theory. Therefore, we expect, based on the optical theorem\cite{Landau1959,Cutkosky1960}, that the absence of the discontinuity for s-wave pairing, as indicated in Eq.~(\ref{D0}), is valid generally. In Appendix~\ref{Appendix3loop} we explicitly analyze the discontinuity up to three-loop order and we present a general procedure to systematically examine the behavior of the discontinuity depending on the superconducting gap symmetry in arbitrary order perturbation theory.

Hereafter, we discuss the sign-changing pairing $\Delta_{\mathbf k_F+\mathbf Q}=-\Delta_{\mathbf k_F}$. In the weak coupling case $\omega_{\text{sf}} \gg \Delta $ ($\lambda \ll 1$), the characteristic scale of the internal bosonic propagator in the continuum region is $\chi_{\mathbf Q}(\omega) = r- \Pi_{\mathbf Q}^{(2)}(\omega) = r- \gamma \Delta f(\frac \omega \Delta) \sim r$, which leads to a non-critical $g^4$ dependence for the discontinuity corrections.  Furthermore, the spectral weight of the resonance is exponentially suppressed as can be seen in (\ref{r3}), therefore the $\delta D$ contributions from the resonance region of the internal bosonic line are small compared to $D_0$. Thus, a systematic expansion in the coupling parameter $g$ is justified in the weak coupling regime  $\omega_{\text{sf}} \gg \Delta$, such that  vertex corrections to the discontinuity $|\delta D| \ll D_0$ are suppressed by higher powers of the coupling parameter $g$. As stressed earlier, in this  limit the magnetic correlation length $\xi \sim 1/\sqrt{r}$ is small and our continuum theory is not the appropriate starting point.

\begin{figure}
\centering
\includegraphics[width=0.4\textwidth]{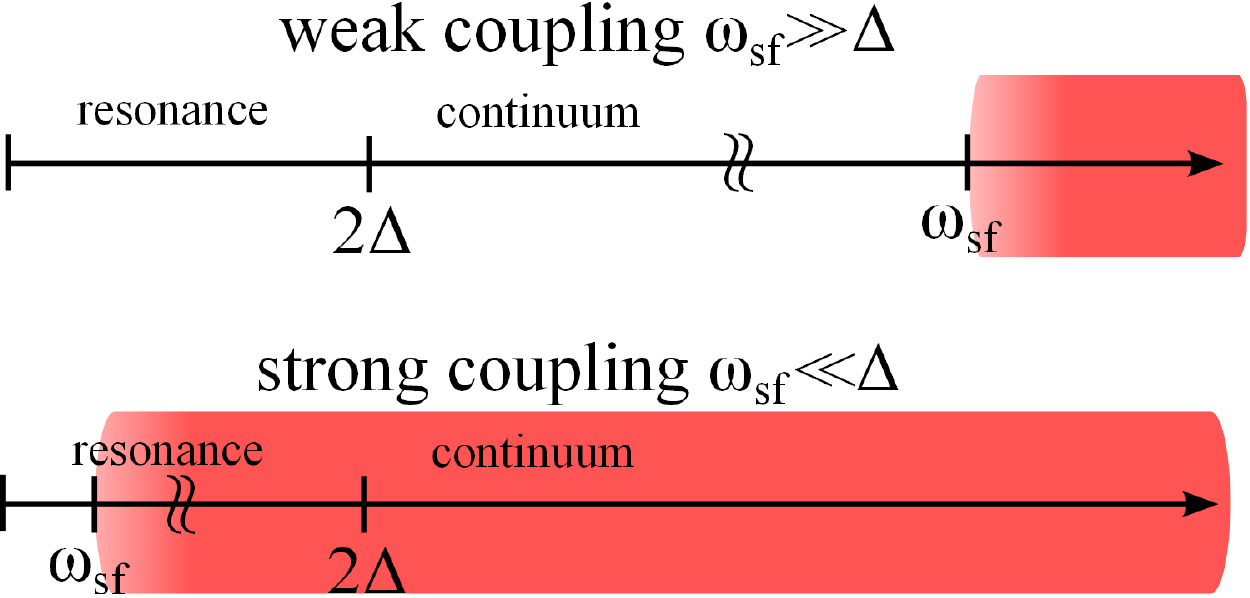}
\caption{Comparison of  energy scales in the superconducting state. In the weak coupling limit $\omega_{\text{sf}} \gg \Delta$ the quantum critical contributions above $\omega_{\text{sf}}$ are not important for the low-energy resonance $\Omega_{\text{res}}<2 \Delta$. In the strong coupling limit quantum critical excitations $\omega>\omega_{\text{sf}}$ determine the bosonic spectrum.}
\label{weakvsstrongcoupling}
\end{figure}

\begin{figure}
\centering
\includegraphics[width=0.47\textwidth]{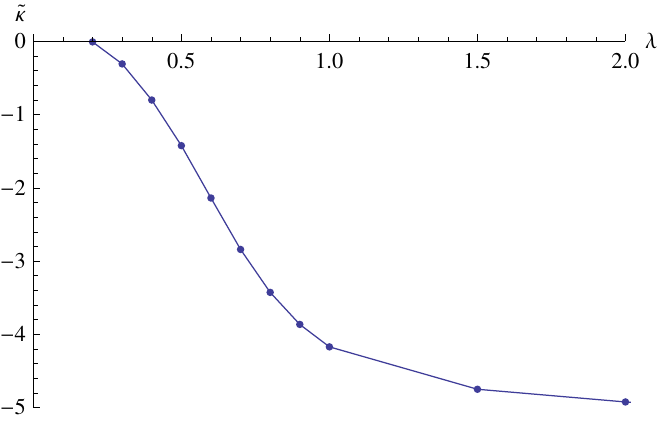}
\caption{Numerical function $\tilde \kappa(\lambda)$ using the calculated  spin susceptibility in the case of $N=8$ hot spots. Details of the analysis are shown in the Appendix~\ref{appendixepsilon}.}
\label{kappa}
\end{figure}

In the strong coupling regime $\omega_{\text{sf}} \ll \Delta$ ($\lambda \gg 1$), we show in Appendix~\ref{appendix3} [see Eq.~(\ref{appendix3.d.6})] that the  the discontinuity correction can be written in the form
\begin{align}
\delta D &= \frac{D_0}{N} \kappa(\frac{\omega_{\text{sf}}}\Delta, \hat \Delta) ,
\end{align}
where the $g$-dependence is confined to the dimensionless function $\kappa(\frac{\omega_{\text{sf}}}\Delta, \hat \Delta)$. Here, we defined the dimensionless parameter
\begin{align}
\hat \Delta = \frac{c_s \Delta}{v_F^2 \gamma} \simeq \frac{c_s T_c}{v_F^2 \gamma} = \frac{9}{N^2} C_2(\lambda) .
\end{align}
At the same time, $\frac{\omega_{\text{sf}}}{\Delta} \simeq \frac{\omega_{\text{sf}}}{T_c} = \frac{1}{4 \lambda^2 C_2(\lambda)}$, therefore the function $\kappa$ depends only on the coupling constant $\lambda$ and
\begin{align}
\delta D &= \frac{D_0}{N} \tilde \kappa(\lambda)  .
 \end{align} 
 The function $\tilde \kappa$ can be computed numerically and is shown in Fig.~\ref{kappa} for $N=8$. It vanishes for small $\lambda \ll 1$. Thus in the weak coupling regime vertex corrections are suppressed by a higher power in the perturbative parameter $g$ and the one-loop calculation is controlled. In the other limit $\lambda \gg 1$ the function is constant $\tilde \kappa(\infty) \approx -5.3$. Therefore, $\delta D \sim D_0$ due to quantum-critical spin fluctuations at energy scales $\omega> \omega_{\text{sf}}$. Technically, in the strong coupling limit these fluctuations determine both the resonance and the continuum region $\omega> 2\Delta$ in Eq.~(\ref{smpapervc3}).   The corresponding scales in the weak and strong coupling regimes are displayed in Fig.~\ref{weakvsstrongcoupling}, where red regions display the quantum-critical contributions. The vertex corrections to the resonance mode are dominated by the continuum region, where the physics is similar to that of the normal state. Therefore, it is not obvious that a $1/N$-expansion of those corrections is permissible. In addition, the numerical values of $\delta D$ are not small, even for the physically relevant case of $N=8$ hot spots. Thus, we conclude that the perturbation expansion  is not controlled in $d=2$.

\subsubsection*{Systematic expansion in $\protect\varepsilon =3-d$ for strong coupling limit}

The above results show that unless $\lambda \ll 1$ no controlled perturbative  expansion exists in $d=2$. Formally the higher order corrections are of order $1/N$ with $N$ the number of fermion species, yet previous results for $d=2$ demonstrated that the standard loop-expansion can not be understood as an expansion in $1/N$\, ~\cite{SSLee2009}. To avoid this problem (albeit in a different context) Mross et al.~\cite{Mross2010} suggested in case of a related problem to combine the $1/N$-expansion  with a further expansion in the parameter $z_{b}-z_{b}^{\ast}$, where $z_{b}$ is the dynamical critical exponent of the boson field and $z_{b}^{\ast }$ is the value of $z_{b}$ where quantum-critical corrections become logarithmic. For the problem of fermions coupled to a fluctuating transverse gauge field, discussed in Ref.~\onlinecite{Mross2010}, $z_{b}^{\ast }=2$. Adapting this approach to our problem yields $z_{b}^{\ast}=1$. However, such an approach is somewhat problematic for the determination of the resonance mode as the boson-dynamics is supposed to be the result of the calculation, i.e. we want to determine the relevant value of $z_{b}$. If indeed the running boson propagator that determines $D$ would be governed by the resonance mode itself, we would have a consistent theory, as the resonance mode is indeed characterized by a dynamic scaling exponent $z_{b}=1 $. However, our results for $d=2$ clearly demonstrate that higher order corrections to the resonance mode have their origin in normal-state quantum-critical excitations with $z_{b}=2$. On the other hand, our result~(\ref{self energy d}) shows that the upper critical dimension of the quantum-critical behavior is $d_{\text{uc}}=3$. Therefore, we propose to use the $\varepsilon$-expansion~(\ref{epsexp}) instead of the expansion of Ref.~\onlinecite{Mross2010}.

The leading order vertex correction at the hot spots in the strong coupling limit $\lambda \rightarrow \infty$ is given by
\begin{align}
\delta \Gamma &(i \Omega, i \omega) = \begin{matrix} \includegraphics[width=3cm]{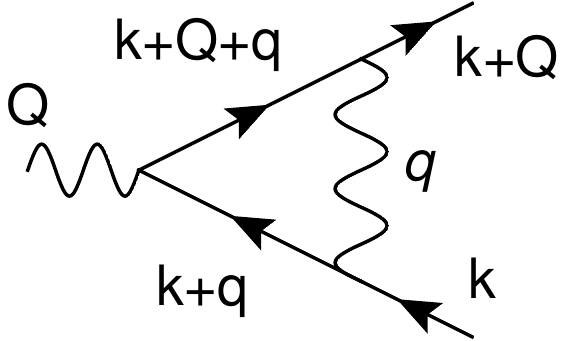}   \end{matrix} ,      \label{vertexpic}
\end{align}
where $q=(i \nu, \mathbf q	)$, $k=(i \Omega, \mathbf{k_F})$ and $Q=(i \omega, \mathbf Q)$. As only the bosonic field depends on the $z$ component of the momentum (assuming fermions are restricted to the two-dimensional xy-plane) we can integrate the bosonic propagator over $q_z$. As a result we find (see Appendix~\ref{appendixepsilon} for details)
\begin{align}
\delta \Gamma& \sim g^{\frac{2}{\epsilon}-1} \int dx \, dy \, d \tilde \nu  \biggl( \frac{1}{\hat \Delta(x^2+y^2) - \frac{\Pi_{\mathbf Q}(i \tilde \nu \Delta)}{\gamma \Delta}}   \biggr)^{\frac{1+\epsilon}{2}}  \nonumber \\
& \hspace*{2cm}\times \,  f(x,y, i \tilde \Omega, i \tilde \omega)  ,   \label{vertexpic2}
\end{align}
where $x=\varepsilon_{\mathbf{k_F+q}}/\Delta$, $y=\varepsilon_{\mathbf{k_F+Q+q}}/\Delta$, $\tilde \nu=\nu/ \Delta$ and $\tilde \Omega = \Omega/\Delta, \tilde \omega = \omega /\Delta$ are the external frequencies in units of the superconducting gap. In the superconducting state the  polarization operator on the imaginary axis can be split into  the resonance contribution  (for $|\omega|<2 \Delta$)  where $\Pi_{\mathbf Q}(\omega) \sim \gamma \omega^2 /\Delta$ and  the continuum contribution (for $|\omega|>2 \Delta$) with $\Pi_{\mathbf Q}(\omega) \sim \gamma \omega$, see Ref.~\onlinecite{Abanov2001}. In both cases, the ratio $\Pi_{\mathbf{Q}}(i   \tilde \nu \Delta)/\gamma \Delta $ is independent of the coupling constant $g$, confining the $g$ dependence of the integral in~(\ref{vertexpic2}) to the parameter $\hat \Delta \sim g^{\frac{4-4\epsilon}{\epsilon}}$. This term dominates the behavior of the bosonic propagator if 
$$ \hat \Delta(x^2+y^2) >  \frac{\Pi_{\mathbf Q}(i \tilde \nu \Delta)}{\gamma \Delta}\, .$$
Since this condition only holds for a small region of the phase space, we can show that these potentially critical contributions yield even higher powers of $g$  for $d>2$.  

The above analysis yields for the correction to the discontinuity 
\begin{equation}
\frac{D-D_{0}}{D_{0}}\propto \gamma ^{\alpha }
\end{equation}%
with the exponent 
\begin{equation}
\alpha =\frac{2}{\protect\varepsilon}-2>0.
\end{equation}%
 The result agrees with our previous findings for $d=2$~(i.e. for $\varepsilon =1$ where the exponent $\alpha $ vanishes). However, for small $\varepsilon $ the exponent $\alpha $ is large, such that vertex corrections are small (exactly for $d=3$ they are exponentially small). A similar treatment can be performed for the self-energy corrections, which is not of importance as they can be controlled by a large $N$ theory. Thus, while the resonance mode for $d=2$ cannot be determined in a controlled fashion, systems with three-dimensional spin excitation spectrum can be well described in terms of the weak coupling theory.

\section{Summary}
In summary, we investigated the role of higher order corrections for the theoretical description of the resonance mode within the spin fermion model. Within this model the occurrence of a resonance mode can be traced back to the emergence of a discontinuity in the imaginary part of the bosonic, spin-self energy at twice the gap value $\Delta$ at the hot spots of the Fermi surface. 

First, we explicitly show that even if one includes higher order corrections the resonance mode only emerges if the phase of the superconducting gap function $\Delta_{\mathbf k_{F}}$ and $\Delta_{\mathbf k_{F}+\mathbf Q}$ are distinct. Thus, we expect the one-loop result 
\begin{align}
D\sim\sin^{2}\left(\frac{\phi_{1}-\phi_{2}}{2}\right)
\end{align}
for the height of the discontinuity to be valid more generally. Here, $\phi_{1}$ and $\phi_{2}$ are the phases of the superconducting order parameter at hot spots connected by the magnetic ordering vector $\mathbf{Q}$. This behavior makes the resonance mode a powerful tool to investigate the inner structure of the pairing condensate.

Second, we find for the two-loop vertex and self-energy correction to $D$ the following result
\begin{align}
\frac{\delta D}{D_0} = \frac{g^2}{N} \cdot f(g),
\end{align}
where $D_0$ is the one-loop  result. The function $f(g)$ is determined by the boson dynamics.  In the strong coupling limit $\lambda \gg 1$ (or $\xi = \sqrt{c_s/r} \gg \frac{c_s v_F}{g^2}$) and in the case of two dimensions, this function scales as a power law $f(g) \sim 1/g^2$ implying that  in two dimensions the neutron resonance mode in unconventional superconductors is a strong coupling phenomenon. The origin of this behavior is the emergence of  quantum-critical fluctuations at intermediate energies $\omega_{\text{sf}} \lesssim \omega \sim \Delta$ that dominate the low energy spin dynamics in the superconducting state. While such quantum critical fluctuations occur for all dimensions $d\leq 3$, they can be analyzed in a controlled fashion by means of the $\varepsilon$-expansion with the small parameter $\varepsilon =3-d$. Quantum critical fluctuations to the resonance mode are now governed by power law behavior, yet the resulting exponents are such that the leading corrections are small. They become of order unity only for $\varepsilon \rightarrow 1$, i.e. for $d=2$. 

Our findings have implications for a number of unconventional superconductors: In case of heavy electron superconductors, such as CeCoIn$_{5}$~\cite{Petrovic2001,Stock2008,Stockert2008}, and the iron-based systems\cite{Christianson2008,Inosov2010}, there are numerous indications that these are three dimensional, albeit moderately anisotropic materials. This implies that we expect the one-loop description of Refs.~[\onlinecite{Abanov2000,Eschrig2000,Abanov2001,Eschrig2002,Abanov2002,Eschrigreview}] to be valid for these materials. The situation is different for the copper-oxide high-temperature superconductors\cite{Fong1999,He2001,He2002} and for the organic charge transfer salts\cite{Malone2010,Lang2003}, that are strongly anisotropic and behave in many ways as two-dimensional systems. Here, our findings imply that a quantitative description of the resonance mode requires going beyond the leading one-loop order. In case of the cuprates, it is tempting to speculate that the observed universal ratio of the resonance mode and the pairing gap in a range of different materials is related to the strong coupling behavior of the resonance mode revealed here. A possible scenario is that higher order corrections modify the one-loop result $\Omega_{\text{res}}/\Delta=f\left(\lambda\right)$ in a way that the function $f\left(\lambda\gg1\right)$ approaches an universal value. In case
of the organic superconductors\cite{Schmalian1998,Kuroki2002}, no neutron measurement of the resonance mode has so far been reported. Numerous experiments support however the existence of an unconventional superconducting state with sign changing order parameter\cite{Malone2010}, i.e. one expects a resonance mode in these systems as well. Our results imply that the highly anisotropic quasi two-dimensional organics should behave similar to the cuprates.

\begin{acknowledgments}
The authors are grateful to A. V. Chubukov and M. Khodas for discussions. This work was supported by the Deutsche Forschungsgemeinschaft through DFG-SPP 1458 ``Hochtemperatursupraleitung in Eisenpniktiden''.
\end{acknowledgments}

\appendix

\begin{widetext}

\section{Self-energies in the superconducting state}   \label{selfenergiesinsc}
To investigate the momentum and frequency dependence of the fermionic self-energy and their influence on the resonance mode, we calculate  the  one-loop self-energy in the superconducting state 
\begin{align}
\Sigma_{\mathbf{k}}^{(p)}(i \Omega_m) &= \includegraphics[width=2cm]{sesc1} = 3 g^2 \frac{T}{L^2} \sum_{\mathbf q, \omega_n}  \chi_{\mathbf{q}} (i \omega_n)  \mathcal G_{\mathbf k+\mathbf q}^{(p)}(i \Omega_m+i \omega_n) ,  \label{ase1}
\end{align}
which we call the propagator self-energy. In a similar fashion we define the hole $\Sigma^{(h)}$ and anomalous self-energy $\Phi$ by replacing $ \mathcal G^{(p)}$ with $ \mathcal G^{(h)}$ or $ \mathcal F$. We can now express the renormalization factors:
\begin{align}
Z_k &= 1- \frac{\Sigma_k ^{(p)} + \Sigma_k^{(h)}}{2 i \omega_n} \quad , \quad \delta \epsilon_k = \frac{\Sigma_k^{(p)}-\Sigma_k^{(h)}}{2} .   \label{Zdeltaepsilon}
\end{align}
After performing the Matsubara sum, we find for the imaginary part of the one-loop particle self-energy on the real axis
\begin{align}
\text{Im}\Sigma_{\mathbf{k}}^{(p)}(\Omega) &= - \frac{3g^2}{L^2} \sum_{\mathbf q} \biggl[ u_{\mathbf k+\mathbf Q}^2 \text{Im}\chi_{\mathbf{q}}(\Omega-\xi_{\mathbf{k}+\mathbf{q}})  \theta(\Omega-\xi_{\mathbf{k}+\mathbf{q}}) + v_{\mathbf k+\mathbf Q}^2 \text{Im} \chi_{\mathbf{q}}(-\Omega-\xi_{\mathbf{k}+\mathbf{q}})  \theta(-\Omega-\xi_{\mathbf{k}+\mathbf{q}})   \biggr]  . \label{ase2}
\end{align}
 We  see that the imaginary part of the normal self-energy is zero for $|\Omega|<\Delta+\Omega_{\text{res}}$ due to the gapped spectrum of both the resonance mode and the fermionic quasiparticles at the hot spots. The same holds for the anomalous self-energy. For $\lambda \leq 1$ this means that the fermionic excitations around $\Omega \approx \Delta$ are well defined. Therefore, we only have to analyze the real part of the self-energies if we are interested in their influence to the discontinuity. Using the Kramers-Kronig relation and the asymmetry of $\text{Im} \chi$ in frequency we find for the real part of the self-energy
\begin{align}
\text{Re}\Sigma_{\mathbf{k}}^{(p)}(\Omega)  &=- \frac{3g^2}{L^2} \sum_{\mathbf q}  \int_{\xi_{\mathbf{k}+\mathbf{q}}}^{\infty} \frac{d\epsilon}{\pi} \text{Im} \chi_{\mathbf{q}}(\epsilon-\xi_{\mathbf{k}+\mathbf{q}})  \frac{(u_{\mathbf{k}+\mathbf{q}}^2-v_{\mathbf{k}+\mathbf{q}}^2)\epsilon+(u_{\mathbf{k}+\mathbf{q}}^2+v_{\mathbf{k}+\mathbf{q}}^2)\Omega}{\epsilon^2-\Omega^2}   \label{ase3} \\
&=- \frac{3g^2}{L^2} \sum_{\mathbf q}  \int_{0}^{\infty} \frac{d\epsilon}{\pi} \text{Im}\chi_{\mathbf{q}}(\epsilon)  \frac{\Omega+ \frac{\varepsilon_{\mathbf{k}+\mathbf{q}}}{\xi_{\mathbf{k}+\mathbf{q}}} \epsilon+ \varepsilon_{\mathbf{k}+\mathbf{q}}}{(\epsilon+\xi_{\mathbf{k}+\mathbf{q}})^2-\Omega^2}  . \nonumber
\end{align}
Since we want to consider small derivations from the hot spots $\delta \mathbf{k}=\mathbf{k} -\mathbf{k_F}$  and from the AF wave vector $\mathbf{p}=\mathbf{q}-\mathbf{Q}$,  we expand
$$ \varepsilon_{\mathbf{k}+\mathbf{q}} = \varepsilon_{\mathbf{k_F}+\mathbf{Q}+\mathbf{p}+\delta \mathbf{k}}  \approx \mathbf{v}_{\mathbf{k}_F+\mathbf{Q}} \cdot (\mathbf{p}+\delta \mathbf{k}) = v_F p_\perp + \varepsilon_{\mathbf{k}+\mathbf{Q}},$$
where $p_\perp$ is the  component of $\mathbf{p}$ perpendicular to the Fermi surface. We introduce dimensionless variables $x=v_F p_\perp/\Delta, y= v_F p_\parallel/\Delta, z=\epsilon/\Delta, \delta=\varepsilon_{\mathbf{k}+\mathbf{Q}}/\Delta, \tilde \Omega=\omega/\Delta$ and the dimensionless RPA spin susceptibility
\begin{align}
\tilde \chi_{x,y}(z) &= \gamma \Delta \chi_{\mathbf q} (z \cdot \Delta)\biggr|_{(\mathbf q- \mathbf Q)^2 = \frac{\Delta^2}{v_F^2}(x^2+y^2)} = \frac{ \gamma \Delta }{r+ c_s \frac{\Delta^2}{v_F^2} (x^2+y^2)- \Pi_{\mathbf Q}^{(2)}(\nu \cdot \Delta)} = \frac{1}{\frac{\omega_{\text{sf}}}\Delta+ \hat \Delta  (x^2+y^2) -\frac{\Pi_{\mathbf Q}^{(2)}(\nu \cdot \Delta)}{\gamma \Delta}}    \, . \label{ase4}
\end{align}
Here, we defined the dimensionless pairing parameter $\hat \Delta = \frac{c_s \Delta}{v_F^2 \gamma}$.  Finally we can express the self-energies as
\begin{align}
\begin{split}
{\text{Re}}\Sigma_{\mathbf{k}}^{(p/h)}(\Omega)  &={\text{Re}}\Sigma_{\mathbf{k}}^{(+/-)}(\Omega) =- \frac{3 \Delta}{2 \pi^2 N} \int dx dy \int_0^\infty dz {\text{Im}} \tilde \chi_{x,y}(z) \frac{\tilde \Omega \pm \frac{x+\delta}{(x+\delta)^2+1} \cdot z  \pm  x \pm  \delta}{(z+\sqrt{(x+\delta)^2+1})^2-\tilde \Omega^2}     \\
\text{Re}\Phi_{\mathbf{k}}(\Omega)  &= \frac{3 \Delta}{2 \pi^2 N} \int dx dy \int_0^\infty dz \text{Im}\tilde \chi_{x,y}(z) \frac{ \frac{z}{\sqrt{(x+\delta)^2+1}}+1}{(z+\sqrt{(x+\delta)^2+1})^2-\tilde \Omega^2}   \\
\end{split}   \label{ase5}
\end{align}
where $\Delta =|\Delta|$ is assumed to be real, which we can locally choose at the hot spot. Note that the anomalous self-energy $\Phi_k$ switches sign for $\mathbf k \rightarrow \mathbf{k+Q}$ just like the superconducting gap. We approximate the spin susceptibility for positive $z$ as
\begin{align}
\text{Im} \tilde \chi_{x,y}(z) = 2 \pi e^{-\frac{\omega_{\text{sf}}}\Delta- \hat \Delta  (x^2+y^2)} \delta(z-\tilde \Omega_{\text{res}}) + \frac{z}{[\frac{\omega_{\text{sf}}}\Delta+ \hat \Delta  (x^2+y^2)]^2+z^2} \theta(z-2), \label{ase6} 
\end{align}
where the first term describes the resonance at $\tilde \Omega_{\text{res}}= 2 (1-e^{-\frac{\omega_{\text{sf}}}\Delta- \hat \Delta  (x^2+y^2)})$ and the second term uses the normal state behavior to express the continuum region. Using $\Delta \approx T_c$ it is possible to express (here $d=2$)
\begin{align}
\frac{\omega_{\text{sf}}}{\Delta} &= \frac{1}{4 \lambda^2 C_2(\lambda)}    \qquad \qquad \hat \Delta = \frac{9}{N^2} C_2(\lambda)    \label{ase7}
\end{align}
 depending only on the dimensionless coupling constant $\lambda$ and the function $C_2(\lambda)$ defined in~(\ref{Tc}). Expanding the formulas~(\ref{ase5}) around $\tilde \Omega=\pm 1$ and $\delta \approx 0$ it is possible to determine numerically the coefficients in
\begin{align}
\begin{split}
Z_{\mathbf k}(\Omega) &= 1 - \frac{\text{Re}\Sigma_{\mathbf{k}}^{(p)}(\Omega)+\text{Re}\Sigma_{\mathbf{k}}^{(h)}(\Omega)}{2 \Omega} \approx Z_0 + Z_f (|\Omega|-\Delta) + Z_m \varepsilon_{\mathbf k+\mathbf Q}^2 ,  \\
\delta \varepsilon_{\mathbf k}(\Omega) &=  \frac{\text{Re}\Sigma_{\mathbf{k}}^{(p)}(\Omega)-\text{Re}\Sigma_{\mathbf{k}}^{(h)}(\Omega)}{2 } \approx  \nu_m  \varepsilon_{\mathbf k+\mathbf Q} + \mathcal{O}(\varepsilon_{\mathbf k+\mathbf Q}^2, (|\Omega|-\Delta) \cdot \varepsilon_{\mathbf k+\mathbf Q}),  \\
\Phi_{\mathbf k}(\Omega) & \approx \Phi_0 + \Phi_f (|\Omega|-\Delta) + \Phi_m \varepsilon_{\mathbf k+\mathbf Q}^2,   \\
\Delta_{\mathbf k}(\Omega) &= \frac{\Phi_{\mathbf k}(\Omega)}{Z_{\mathbf k}(\Omega)} \approx \underbrace{\frac{\Phi_0}{Z_0}}_{\Delta_0 \approx \Delta} + \underbrace{\frac{Z_0 \Phi_f-\Phi_0 Z_f}{Z_0^2} }_{\Delta_f}(|\Omega|-\Delta) +  \underbrace{\frac{Z_0 \Phi_m-\Phi_0 Z_m}{Z_0^2} }_{\Delta_m}\varepsilon_{\mathbf k+\mathbf Q}^2 .
\end{split}   \label{expans}
\end{align}
For $\lambda \lesssim 1$ all parameters are small compared to $1$ and the expansion is a good approximation around the hot spots, but for $\lambda \gg 1$ the frequency parameters $Z_f, \Phi_f, \Delta_f$ become large, because the resonance energy $\Omega_{\text{res}} \ll \Delta$ is small compared to the gap and the real part of the self-energies at $\omega=\Omega_{\text{res}}+\Delta \approx \Delta$ develops a resonance near the expansion region.

\section{Self-energies corrections for the discontinuity}
Using~(\ref{impi12}) and $\text{Im}\mathcal G_{\mathbf k}(|\omega|<\Delta')=0$ the imaginary part of the polarization operator for zero temperature including the dressed propagators can be written as
 \begin{align}
\text{Im} \Pi_{\mathbf{q}}^{(2)}(\omega) &=  \frac{2 g^2}{L^2} \sum_{\mathbf k}  \int_{\Delta'}^{\omega-\Delta'} \frac{d\lambda}{\pi}     \biggl[ {\rm Im} {\cal G}_{\mathbf{k}}^{(h)}\left( \lambda \right) {\rm Im} %
{\cal G}_{\mathbf{k+q}}^{(h)}\left( \lambda+\omega \right) + {\rm Im} {\cal F}_{\mathbf{k}}\left( \lambda \right) {\rm Im} {\cal F}_{%
\mathbf{k+q}}^{* }\left( \lambda +\omega \right) \biggr] ,   \label{appendixsecorr1}
 \end{align}
where $\Delta' \lesssim \Delta$ is the minimal excitation energy around the hot spot. Inserting the imaginary parts of the propagator we find for the $\mathcal G \mathcal G$ contribution for $\omega \approx 2 \Delta$ and external momentum $\mathbf{Q}$
\begin{align}
\text{Im} \Pi_{\mathcal G \mathcal G,\mathbf{Q}}^{(2)}(\omega) &=  \frac{2 g^2 \pi }{L^2} \sum_{\mathbf k} \int_{\Delta'}^{\omega-\Delta'}  d\lambda  \frac{v_{\mathbf{k}}(\lambda)^2 u_{\mathbf{k+Q}}(\lambda-\omega)^2}{Z_{\mathbf{k}}(\lambda)Z_{\mathbf{k+Q}}(\lambda-\omega)} \delta \biggl( \lambda- \sqrt{ \bigl[\frac{\varepsilon_{\mathbf k}+\delta\varepsilon_{\mathbf k} }{Z_{\mathbf{k}}(\lambda)} \bigr]^2 +\Delta_{\mathbf{k}}(\lambda)^2}  \biggr)  \, \times \\
& \hspace{5cm}  \delta \biggl( \lambda-\omega + \sqrt{ \bigl[\frac{\varepsilon_{\mathbf {k+Q}}+\delta\varepsilon_{\mathbf {k+Q}} }{Z_{\mathbf{k+Q}}(\lambda-\omega)} \bigr]^2 +\Delta_{\mathbf{k+Q}}(\lambda-\omega)^2}  \biggr)
\end{align}
Since the momentum $\mathbf{k} \approx \mathbf{k}_F$ and frequencies $\lambda \approx \Delta, \lambda-\omega \approx - \Delta$ are still restricted around the hot spots, we perform a frequency expansion in the $\delta$ distributions and use the relations in~(\ref{expans})
\begin{align*}
 \sqrt{ \bigl[\frac{\varepsilon_{\mathbf k}+\delta\varepsilon_{\mathbf k} }{Z_{\mathbf{k}}(\lambda)} \bigr]^2 +\Delta_{\mathbf{k}}(\lambda)^2}  & \stackrel{\lambda \approx \Delta} \approx \sqrt{ \bigl[\frac{\varepsilon_{\mathbf k}+\delta\varepsilon_{\mathbf k} }{Z_{\mathbf{k}}(\Delta)} \bigr]^2 +\Delta_{\mathbf{k}}(\Delta)^2} + \overbrace{\frac{\Delta_f Z_{\mathbf k}(\Delta)^3 \Delta_{\mathbf k}(\Delta)- Z_f (\varepsilon_{\mathbf k}+\delta\varepsilon_{\mathbf k})^2}{Z_{\mathbf k}(\Delta) \sqrt{ \bigl[\frac{\varepsilon_{\mathbf k}+\delta\varepsilon_{\mathbf k} }{Z_{\mathbf{k}}(\Delta)} \bigr]^2 +\Delta_{\mathbf{k}}(\Delta)^2}}}^{\alpha_{\mathbf{k}}} \cdot (\lambda-\Delta) \\
 & \hspace{1.7mm}  = \sqrt{ \bigl[\frac{\varepsilon_{\mathbf k}+\delta\varepsilon_{\mathbf k} }{Z_{\mathbf{k}}(\Delta)} \bigr]^2 +\Delta_{\mathbf{k}}(\Delta)^2} + \alpha_{\mathbf{k}} \cdot (\lambda-\Delta)   \\
 \sqrt{ \bigl[\frac{\varepsilon_{\mathbf {k+Q}}+\delta\varepsilon_{\mathbf {k+Q}} }{Z_{\mathbf{k+Q}}(\lambda-\omega)} \bigr]^2 +\Delta_{\mathbf{k+Q}}(\lambda-\omega)^2}   & \stackrel{\omega-\lambda \atop \approx -\Delta} \approx   \sqrt{ \bigl[\frac{\varepsilon_{\mathbf {k+Q}}+\delta\varepsilon_{\mathbf {k+Q}} }{Z_{\mathbf{k+Q}}(-\Delta)} \bigr]^2 +\Delta_{\mathbf{k+Q}}(-\Delta)^2} - \alpha_{\mathbf{k+Q}} \cdot (\lambda-\omega+\Delta)
 \end{align*} 
Using this relations we write
\begin{align*}
 \delta \biggl( \lambda& - \sqrt{ \bigl[\frac{\varepsilon_{\mathbf k}+\delta\varepsilon_{\mathbf k} }{Z_{\mathbf{k}}(\lambda)} \bigr]^2 +\Delta_{\mathbf{k}}(\lambda)^2}  \biggr) \, \delta \biggl( \lambda-\omega + \sqrt{ \bigl[\frac{\varepsilon_{\mathbf {k+Q}}+\delta\varepsilon_{\mathbf {k+Q}} }{Z_{\mathbf{k+Q}}(\lambda-\omega)} \bigr]^2 +\Delta_{\mathbf{k+Q}}(\lambda-\omega)^2}  \biggr)   \nonumber \\
 &= \frac{\delta(\lambda-\beta_{\mathbf{k}})\delta(\lambda-\omega+\beta_{\mathbf{k+Q}})}{(1-\alpha_{\mathbf{k}})(1-\alpha_{\mathbf{k+Q}})}
\end{align*}
with
\begin{align}
\beta_{\mathbf{k}} = \frac{\sqrt{ \bigl[\frac{\varepsilon_{\mathbf k}+\delta\varepsilon_{\mathbf k} }{Z_{\mathbf{k}}(\Delta)} \bigr]^2 +\Delta_{\mathbf{k}}(\Delta)^2} - \alpha_{\mathbf{k}} \Delta}{1-\alpha_{\mathbf{k}} } .
\end{align}
Evaluating the frequency integration we find
\begin{align}
\text{Im} \Pi_{\mathcal G \mathcal G,\mathbf{Q}}^{(2)}(\omega) &=  \frac{2 g^2 \pi }{L^2} \sum_{\mathbf k}  \frac{v_{\mathbf{k}}(\beta_{\mathbf{k}})^2 u_{\mathbf{k+Q}}(-\beta_{\mathbf{k+Q}})^2}{Z_{\mathbf{k}}(\beta_{\mathbf{k}})Z_{\mathbf{k+Q}}(-\beta_{\mathbf{k+Q}})}  \frac{\delta(\omega-\beta_{\mathbf{k}}-\beta_{\mathbf{k+Q}})}{(1-\alpha_{\mathbf{k}})(1-\alpha_{\mathbf{k+Q}})}
\end{align}
Now, we expand $\beta_{\mathbf{k}}$ till second order in $\varepsilon_{\mathbf{k}},\varepsilon_{\mathbf{k+Q}} \ll \Delta$ 
\begin{align}
\beta_{\mathbf{k}}  \approx \Delta + \frac{(\varepsilon_{\mathbf{k}}+\nu_m \varepsilon_{\mathbf{k+Q}})^2}{2 \Delta Z_0^2 (1- \Delta_f)} + \frac{\Delta_m \varepsilon_{\mathbf{k+Q}}^2}{1-\Delta_f} .
\end{align}
In the considered regime $\lambda \leq 1$ the last term is negligible, because $| \Delta_m Z_0^2 \Delta| \ll 1 , |\nu_m|$ . We see that the minimal excitation energy $\Delta '=\Delta$ remains the same even including self-energy corrections and the discontinuity still appears at $2 \Delta$. Also the momentum contributing to the discontinuity $\varepsilon_{\mathbf{k}}=\varepsilon_{\mathbf{k+Q}} =0$ are again restricted to the hot spots. After the usual linearization $\epsilon=\varepsilon_{\mathbf{k}}, \epsilon'=\varepsilon_{\mathbf{k+Q}}$ we substitute $x=(\epsilon+\nu_m \epsilon')/\sqrt{2 \Delta Z_0^2 (1-\Delta_f)}$ and $y=(\epsilon'+\nu_m \epsilon)/\sqrt{2 \Delta Z_0^2 (1-\Delta_f)}$ and find for the $\mathcal G \mathcal G$ contribution of the discontinuity 
\begin{align}
D_{\mathcal G \mathcal G} &= \lim_{\delta \rightarrow 0^+} \text{Im} \Pi_{\mathcal G \mathcal G,\mathbf{Q}}^{(2)}(2 \Delta+\delta)    \nonumber \\
&= \frac{g^2 N}{2 \pi v_F^2}  \frac{v_{\mathbf{k}}(\beta_{\mathbf{k}})^2 u_{\mathbf{k+Q}}(-\beta_{\mathbf{k+Q}})^2}{Z_{\mathbf{k}}(\beta_{\mathbf{k}})Z_{\mathbf{k+Q}}(-\beta_{\mathbf{k+Q}})(1-\alpha_{\mathbf{k}})(1-\alpha_{\mathbf{k+Q}})}   \biggr|_{\varepsilon_{\mathbf{k}}=0 \atop \varepsilon_{\mathbf{k+Q}} =0}    \times \nonumber \\
& \hspace{2cm}  \frac{2 \Delta Z_0^2 (1-\Delta_f)}{1-\nu_m^2}   \underbrace{\lim_{\delta \rightarrow 0^+} \int dx \, dy \,  \delta(\delta-x^2-y^2) }_{\pi}  \nonumber \\
&=  \frac{D_0}{2(1-\Delta_f)(1-\nu_m^2)}
\end{align}
The anomalous $\mathcal F \mathcal F^*$ contribution gives the same contribution, but depending on the gap symmetry we find
\begin{align}
D_{\mathcal F \mathcal F^*} = \frac{D_0}{2(1-\nu_m^2)(1-\Delta_f)}  \text{ sign}( -\Delta_{\mathbf{k}_F}/\Delta_{\mathbf{k}_F+\mathbf{Q}}) .
\end{align}
Therefore, the discontinuity vanishes again for the s-wave symmetry $\Delta_{\mathbf{k}_F}=\Delta_{\mathbf{k}_F+\mathbf{Q}}$.

\section{Evaluation of the Matsubara summation} \label{appendix2}
In order to get the correction $\delta D$ to the discontinuity we have to calculate the imaginary part of $\delta \Pi_{\mathbf Q}(\omega)$. Thus, we have to execute and analytically continue the double Matsubara summation
\begin{align}
Q_{\mathbf k,\mathbf q}(i \omega_n) &= T^2 \sum_{\Omega_m, \nu_k} A_{\mathbf k}( i \Omega_m) B_{\mathbf k+\mathbf Q}(i \Omega_m+i \omega_n) C_{\mathbf k+\mathbf q+\mathbf Q}(i \Omega_m+ i \omega_n + i \nu_k)   D_{\mathbf k+\mathbf q}(i \Omega_m+ i \nu_k) \chi(i \nu_k) ,   \label{matssum1}
\end{align}
where $A,B,C,D$ are the different combinations of fermionic propagators in the superconducting state. For $T=0$ the fermionic $f(\epsilon)=\theta(-\epsilon)$ and bosonic distributions functions $g(\epsilon)=-g(-\epsilon)$ severaly restrict the phase space for the considered case $\omega>0$. Using the identities for the fermionic Green's functions
\begin{align}
\begin{split}
A_{\mathbf k}^{R/A}(x) &=\frac{u_{A,\mathbf k}^2}{x-\xi_{\mathbf k}\pm i0} +  \frac{v_{A,\mathbf k}^2}{x+\xi_{\mathbf k}\pm i 0}    , \\
\text{Im }  A_{\mathbf k}^{R/A}(x) &= \mp \pi [ u_{A,\mathbf k}^2 \delta(x-\xi_{\mathbf k}) + v_{A,\mathbf k}^2 \delta(x+ \xi_{\mathbf k}) ]  , \\
 A_{\mathbf k} (x) &= \text{Re}  A_{\mathbf k}^{R/A}(x) = \mathcal P \frac{u_{A,\mathbf k}^2}{x-\xi_{\mathbf k}} + \mathcal P \frac{v_{A,\mathbf k}^2}{x+\xi_{\mathbf k}}    ,
\end{split}  \label{appendix2.6}
\end{align}
it can be shown that the analytical continuation of (\ref{matssum1}) yields
\begin{align}
 \text{Im}  Q(\omega) &=  v_{A,\mathbf k}^2 u_{B,\mathbf k+\mathbf Q}^2 \delta(\omega- \xi_\mathbf k -\xi_{\mathbf k+\mathbf Q})    \int_0^\infty dx\,   \text{Im} \chi_{ \mathbf q}^R(-x) C_{\mathbf k+\mathbf q+\mathbf Q}(-x+\xi_{\mathbf k+\mathbf Q}) D_{\mathbf k+\mathbf q}(-x-\xi_\mathbf k) \nonumber   \\
& \quad - u_{C,\mathbf k+\mathbf q+\mathbf Q}^2  v_{D,\mathbf k+\mathbf q}^2  \delta(\omega-\xi_{\mathbf k+\mathbf q+\mathbf Q}-\xi_{\mathbf k+\mathbf q}) \int_0^\infty dx\,   \text{Im}  \chi_{ \mathbf q}^R(x) A_\mathbf k(-x-\xi_{\mathbf k+\mathbf q}) B_{\mathbf k+\mathbf Q}(-x+\xi_{\mathbf k+\mathbf q+\mathbf Q})      \nonumber \\
& \quad +  v_{A,\mathbf k}^2  u_{C,\mathbf k+\mathbf q+\mathbf Q}^2   \text{Im}  \chi_{ \mathbf q}^R(-\omega+\xi_\mathbf k+\xi_{\mathbf k+\mathbf q+\mathbf Q}) B_{\mathbf k+\mathbf Q}( \omega-\xi_\mathbf k ) D_{\mathbf k+\mathbf q}(-\omega+\xi_{\mathbf k+\mathbf q+\mathbf Q}  ) \theta(\omega-\xi_\mathbf k-\xi_{\mathbf k+\mathbf q+\mathbf Q}) \nonumber \\
& \quad -u_{B,\mathbf k+\mathbf Q}^2 v_{D,\mathbf k+\mathbf q}^2   \text{Im}  \chi_{\mathbf q}^R(\omega-\xi_{\mathbf k+\mathbf q}-\xi_{\mathbf k+\mathbf Q})  A_\mathbf k(-\omega+\xi_{\mathbf k+\mathbf Q})  C_{\mathbf k+\mathbf q+\mathbf Q}(\omega-\xi_{\mathbf k+\mathbf q})   \theta(\omega-\xi_{\mathbf k+\mathbf q}-\xi_{\mathbf k+\mathbf Q})  \nonumber  \\
& \quad -\pi  v_{B,\mathbf k+\mathbf Q}^2  u_{C,\mathbf k+\mathbf q+\mathbf Q}^2  v_{D,\mathbf k+\mathbf q}^2 A_\mathbf k (-\omega-\xi_{\mathbf k+\mathbf Q})  \chi_{ \mathbf q}(\omega+\xi_{\mathbf k+\mathbf Q}-\xi_{\mathbf k+\mathbf q} ) \delta(  \omega -\xi_{\mathbf k+\mathbf q} - \xi_{\mathbf k+\mathbf q+\mathbf Q} )    \nonumber \\
& \quad  - \pi v_{A,\mathbf k}^2  u_{C,\mathbf k+\mathbf q+\mathbf Q}^2 v_{D,\mathbf k+\mathbf q}^2  B_{\mathbf k+\mathbf Q}( \omega-\xi_\mathbf k )  \chi_{ \mathbf q}(\xi_\mathbf k-\xi_{\mathbf k+\mathbf q} )    \delta(  \omega -\xi_{\mathbf k+\mathbf q} - \xi_{\mathbf k+\mathbf q+\mathbf Q} )  \nonumber \\
& \quad  -\pi   v_{A,\mathbf k}^2 u_{B,\mathbf k+\mathbf Q}^2 v_{D,\mathbf k+\mathbf q}^2 C_{\mathbf k+\mathbf q+\mathbf Q}(\omega-\xi_{\mathbf k+\mathbf q}) \chi_{ \mathbf q}(\xi_\mathbf k-\xi_{\mathbf k+\mathbf q}) \delta(\omega- \xi_\mathbf k -\xi_{\mathbf k+\mathbf Q}) \nonumber \\
& \quad -\pi  v_{A,\mathbf k}^2 u_{B,\mathbf k+\mathbf Q}^2  v_{C,\mathbf k+\mathbf q+\mathbf Q}^2 D_{\mathbf k+\mathbf q}(-\omega-\xi_{\mathbf k+\mathbf q+\mathbf Q}) \chi_{ \mathbf q}(-\omega+\xi_\mathbf k-\xi_{\mathbf k+\mathbf q+\mathbf Q})  \delta(\omega- \xi_\mathbf k -\xi_{\mathbf k+\mathbf Q})   \nonumber \\
& \quad - \pi^2  v_{A,\mathbf k}^2 u_{B,\mathbf k+\mathbf Q}^2 v_{C,\mathbf k+\mathbf q+\mathbf Q}^2 v_{D,\mathbf k+\mathbf q}^2  \text{Im} \chi_{ \mathbf q}^R(\xi_\mathbf k-\xi_{ k+\mathbf q}) \delta(\omega- \xi_\mathbf k -\xi_{\mathbf k+\mathbf Q}) \delta(\omega+\xi_{\mathbf k+\mathbf q+\mathbf Q}- \xi_{\mathbf k+\mathbf q})  \nonumber \\
& \quad +\pi^2 v_{A,\mathbf k}^2 v_{B,\mathbf k+\mathbf Q}^2 u_{C,\mathbf k+\mathbf q+\mathbf Q}^2 v_{D,\mathbf k+\mathbf q}^2  \text{Im}  \chi_{ \mathbf q}^R(\xi_\mathbf k-\xi_{\mathbf k+\mathbf q}) \delta(\omega- \xi_\mathbf k +\xi_{\mathbf k+\mathbf Q}) \delta(\omega-\xi_{\mathbf k+\mathbf q+\mathbf Q}-\xi_{\mathbf k+\mathbf q})  \, .      \label{appendix2.13}
\end{align}

\section{Evaluation of the momentum-integration for discontinuity from vertex corrections}   \label{appendix3}
The imaginary part of $\Pi_{VC}(\mathbf Q,\omega)$ can be obtained from (\ref{appendix2.13}) by using a similar linearization as explained in Section~(\ref{normalstatesec}):
\begin{align}
\biggl(\frac{1}{L^2}\biggr)^2 \sum_{\mathbf k,\mathbf q \approx \mathbf Q} & f(\varepsilon_{\mathbf k},\varepsilon_{\mathbf k+ \mathbf Q},\varepsilon_{\mathbf k+ \mathbf Q+\mathbf q},\varepsilon_{\mathbf k+ \mathbf q},\Delta_{\mathbf k},\Delta_{\mathbf k+\mathbf Q},\Delta_{\mathbf k+ \mathbf Q+\mathbf q},\Delta_{\mathbf k+ \mathbf q}) g[(\mathbf q-\mathbf Q)^2]\nonumber \\
&= N \biggl( \frac{1}{8 \pi^2 v_\perp v_\parallel} \biggr)^2 \int d\epsilon\, d\epsilon' \,d\lambda \, d\lambda'\,  f(\epsilon,\epsilon',\epsilon+\lambda,\epsilon'+\lambda', \Delta, \pm \Delta, \Delta ,\pm \Delta)  g\biggl[ \biggl( \frac{\lambda+\lambda'}{2 v_\perp} \biggr)^2 + \biggl( \frac{\lambda-\lambda'}{2 v_\parallel} \biggr)^2 \biggr]  \nonumber \\
& \approx N \biggl( \frac{1}{4\pi^2 v_F^2} \biggr)^2 \int d\epsilon\, d\epsilon' \,d\lambda \, d\lambda'\,  f(\epsilon,\epsilon',\epsilon+\lambda,\epsilon'+\lambda', \Delta_{\mathbf k_F}, \pm \Delta_{\mathbf k_F}, \Delta_{\mathbf k_F} ,\pm \Delta_{\mathbf k_F})  g\biggl[ \frac{\lambda^2+\lambda'^2}{ v_F^2} \biggr] \, , \label{appendix1.6}  
\end{align}
describing small derivations from the hot spots by   $ \mathbf p=\mathbf k - \mathbf k_F$ and from the AF wave vector by $\mathbf p'= \mathbf q - \mathbf Q$. Using this approximation we can write
\begin{align}
   \text{Im} \Pi_{VC}(\mathbf Q , \omega) &= -  g^4 N \sum_{\{A,B,C,D\}} \biggl( \frac{1}{8 \pi^2 v_\parallel v_\perp} \biggr)^2 \int d\epsilon \, d\epsilon' \, d\lambda\, d\lambda' \,  \text{Im}   Q(\omega) \, ,    \label{appendix3.1}
  \end{align}  
where we set $\varepsilon_k = \epsilon  , \varepsilon_{k+Q} = \epsilon'  , \varepsilon_{k+q+Q} = \epsilon+\lambda  , \varepsilon_{k+q} = \epsilon'+\lambda'$ and $(\mathbf q-\mathbf Q)^2=\frac{\lambda^2+\lambda'^2}{v_F^2}$ . Later it will be useful to use the following  symmetries of the momentum integration before the linearization of  the spectrum
\begin{align}
\begin{split}
(\mathbf k, \mathbf k+\mathbf Q+ \mathbf q)   &\leftrightarrow (\mathbf k+\mathbf Q, \mathbf k+\mathbf q)   \, ,    \\
 (\mathbf k, \mathbf k+\mathbf Q)   &\leftrightarrow (\mathbf k+\mathbf Q+ \mathbf q, \mathbf k+\mathbf q)      \, .                                              
\end{split}    \label{appendix3.2}
\end{align}
After the linearization of the spectrum the coherence-factors 
\begin{equation*}
\begin{split}
u_{\mathcal G^{(p)},\mathbf{k}}^{2}& =v_{\mathcal G^{(h)},\mathbf{k}}^{2}=\frac{1}{2}\left( 1+%
\frac{\varepsilon _{\mathbf{k}}}{\xi _{\mathbf{k}}}\right) , \\
v_{\mathcal G^{(p)},\mathbf{k}}^{2}& =u_{\mathcal G^{(h)},\mathbf{k}}^{2}=\frac{1}{2}\left( 1-%
\frac{\varepsilon _{\mathbf{k}}}{\xi _{\mathbf{k}}}\right) \\
u_{\mathcal F,\mathbf{k}}^{2}& =-v_{\mathcal F,\mathbf{k}}^{2}=\frac{1}{2}\frac{\Delta _{%
\mathbf{k}}}{\xi _{\mathbf{k}}},
\end{split}%
\end{equation*}
can be written as
\begin{align}
\begin{array}{|c||c|c|c|c|}
\bigotimes& u_{A,\epsilon}^2 = u_{C,\epsilon}^2 & v_{A,\epsilon}^2 = v_{C,\epsilon}^2& u_{B,\epsilon}^2=u_{D,\epsilon}^2  & v_{B,\epsilon}^2=v_{D,\epsilon}^2 \\
 \hline \hline \mathcal G ^{(p)} & \frac{1}{2} \left(1+\frac{\epsilon}{\sqrt{\epsilon^2+|\Delta_{\mathbf k_F}|^2}} \right) &  \frac{1}{2} \left(1-\frac{\epsilon}{\sqrt{\epsilon^2+|\Delta_{\mathbf k_F}|^2}} \right) & \frac{1}{2} \left(1+\frac{\epsilon}{\sqrt{\epsilon^2+|\Delta_{\mathbf k_F}|^2}} \right) &  \frac{1}{2} \left(1-\frac{\epsilon}{\sqrt{\epsilon^2+|\Delta_{\mathbf k_F}|^2}} \right) \\ 
 \hline \mathcal G ^{(h)} & \frac{1}{2} \left(1-\frac{\epsilon}{\sqrt{\epsilon^2+|\Delta_{\mathbf k_F}|^2}} \right) &  \frac{1}{2} \left(1+\frac{\epsilon}{\sqrt{\epsilon^2+|\Delta_{\mathbf k_F}|^2}} \right) & \frac{1}{2} \left(1-\frac{\epsilon}{\sqrt{\epsilon^2+|\Delta_{\mathbf k_F}|^2}} \right) &  \frac{1}{2} \left(1+\frac{\epsilon}{\sqrt{\epsilon^2+|\Delta_{\mathbf k_F}|^2}} \right)  \\
\hline \mathcal F & \frac{1}{2} \frac{\Delta_{\mathbf k_F}}{\sqrt{\epsilon^2+|\Delta_{\mathbf k_F}|^2}}& - \frac{1}{2} \frac{\Delta_{\mathbf k_F}}{\sqrt{\epsilon^2+|\Delta_{\mathbf k_F}|^2}}& \pm \frac{1}{2}\frac{\Delta_{\mathbf k_F}}{\sqrt{\epsilon^2+|\Delta_{\mathbf k_F}|^2}} & \mp \frac{1}{2} \frac{\Delta_{\mathbf k_F}}{\sqrt{\epsilon^2+|\Delta_{\mathbf k_F}|^2}} \\
 \hline \mathcal F^* &\frac{1}{2} \frac{\Delta_{\mathbf k_F}^*}{\sqrt{\epsilon^2+|\Delta_{\mathbf k_F}|^2}}& - \frac{1}{2} \frac{\Delta_{\mathbf k_F}^*}{\sqrt{\epsilon^2+|\Delta_{\mathbf k_F}|^2}}& \pm \frac{1}{2}\frac{\Delta_{\mathbf k_F}^*}{\sqrt{\epsilon^2+|\Delta_{\mathbf k_F}|^2}} & \mp \frac{1}{2} \frac{\Delta_{\mathbf k_F}^*}{\sqrt{\epsilon^2+|\Delta_{\mathbf k_F}|^2}}
 \end{array}    \label{appendix3.3}
\end{align}
where the different signs of the coherence-factors for the anomalous Green's functions occur for the different gap symmetries $\Delta_{\mathbf k_F+\mathbf Q} = \pm \Delta_{\mathbf k_F}$. Because we see from the $\{A,B,C,D\}$ sum, that we have only combinations of $F$ and $F^\dagger$ in the diagrams, there will always occur combinations $\Delta_{\mathbf k_F} \cdot \Delta_{\mathbf k_F}^*= |\Delta_{\mathbf k_F}|^2 =\Delta^2$  and it is allowed to assume $\Delta_{\mathbf k_F}= \Delta_{\mathbf k_F}^*= \Delta $ to be real. Therefore, we can simplify the above table
\begin{align}
\begin{array}{|c||c|c|c|c|c|c|c|c|}
\bigotimes& u_{A,\epsilon}^2  & v_{A,\epsilon}^2 & u_{B,\epsilon}^2  & v_{B,\epsilon}^2 &  u_{C,\epsilon}^2   & v_{C,\epsilon}^2 & u_{D,\epsilon}^2  &v_{D,\epsilon}^2 \\
 \hline \hline \mathcal G ^{(p)} & u_\epsilon^2 & v_\epsilon^2 & u_\epsilon^2 & v_\epsilon^2 & u_\epsilon^2 & v_\epsilon^2 & u_\epsilon^2 & v_\epsilon^2   \\
 \hline \mathcal G ^{(h)} & v_\epsilon^2 & u_\epsilon^2 & v_\epsilon^2 & u_\epsilon^2 & v_\epsilon^2 & u_\epsilon^2 & v_\epsilon^2 & u_\epsilon^2   \\
 \hline \mathcal  F /\mathcal  F^* & u_\epsilon v_\epsilon & - u_\epsilon v_\epsilon & \pm u_\epsilon v_\epsilon & \mp u_\epsilon v_\epsilon & u_\epsilon v_\epsilon & - u_\epsilon v_\epsilon & \pm u_\epsilon v_\epsilon & \mp u_\epsilon v_\epsilon
 \end{array}      \label{appendix3.4}
\end{align}
where we defined using $\xi_\epsilon= \sqrt{\epsilon^2+\Delta^2}$
\begin{align}
\begin{split}
u_\epsilon^2 &= \frac{1}{2} \left(1+\frac{\epsilon}{\sqrt{\epsilon^2+\Delta^2}} \right)=\frac{1}{2} \left(1-\frac{\epsilon}{\xi_\epsilon} \right) \, , \\
 v_\epsilon^2 &= \frac{1}{2} \left(1-\frac{\epsilon}{\sqrt{\epsilon^2+\Delta^2}} \right) =\frac{1}{2} \left(1-\frac{\epsilon}{\xi_\epsilon} \right)  \, .
\end{split}   \label{appendix3.5}
\end{align}
We now have set the framework for the momentum integration and continue in examining the discontinuity of the imaginary part at $\omega = 2 \Delta + \delta$ with $\delta \rightarrow 0$. Because there are four different kinds of terms in $ \text{Im}  Q(\omega)$, we have to analyze them separately. At first we will examine the terms with $\theta$-function, then the terms with $\delta$-function and $x$-integration, thereafter the terms with 3 coherence-factors and at least the terms with 4 coherence-factors. 

\subsection*{Terms with $\theta$-functions}
Using the above linearization we obtain for the first of the terms with the $\theta$-function:
\begin{align}
 \text{Im} \Pi_{VC,\theta,1}(\omega) &= -  g^4 N \biggl( \frac{1}{8 \pi^2 v_\parallel v_\perp} \biggr)^2 \sum_{\{A,B,C,D\}} \int d\epsilon \, d\epsilon' \, d\lambda\, d\lambda' \,   \text{Im}  \chi_{\lambda,\lambda'}( - \omega + \xi_\epsilon + \xi_{\epsilon+\lambda}) v_{A,\epsilon}^2 u_{C,\epsilon+\lambda}^2    \nonumber \\
& \hspace*{5.2cm} B_{\epsilon'}(\omega-\xi_\epsilon)  D_{\epsilon'+\lambda'}(-\omega+\xi_{\epsilon+\lambda}) \theta(\omega-\xi_\epsilon-\xi_{\epsilon+ \lambda})      \, .  \nonumber
\end{align}
Due to the $\theta$-function the phase space of the $(\epsilon,\lambda)$-integration will be restricted strongly at the discontinuity $\omega \rightarrow 2 \Delta + \delta$:
\begin{align}
 \lim_{\delta \rightarrow 0} \int d\epsilon  \,d\lambda   \,\theta(\omega-\xi_\epsilon - \xi_{\epsilon+\lambda}) f(\epsilon,\epsilon+\lambda)&=  \lim_{\delta \rightarrow 0} \int d\epsilon \, d\lambda  \, \theta(2 \Delta + \delta -\xi_\epsilon - \xi_{\epsilon+\lambda})f_\delta(\epsilon,\epsilon+\lambda)  \nonumber \\
 & \approx  \lim_{\delta \rightarrow 0}\int d\epsilon \, d\lambda  \,\theta( \delta - \frac{\epsilon^2+(\epsilon+\lambda)^2}{2 \Delta}) f_\delta(\epsilon,\epsilon+\lambda)   \nonumber \\
 &\approx \lim_{\delta \rightarrow 0}\int d\epsilon \, d\lambda  \,\theta( \delta - \frac{\epsilon^2+(\epsilon+\lambda)^2}{2 \Delta}) f_\delta(\sqrt{2 \delta \Delta},\sqrt{2 \delta \Delta})   \nonumber \\
 &=  \lim_{\delta \rightarrow 0} 2 \pi \Delta f_\delta(\sqrt{2 \delta \Delta},\sqrt{2 \delta \Delta}) \cdot \delta   \, .      \label{appendix3.a.1}
 \end{align} 
Thus, these terms will vanish if the remaining function $f_\delta(0,0)$ is non-singular. Using the asymmetry of $ \text{Im}  \chi_{\lambda,\lambda'}(x)= - \text{Im}  \chi_{\lambda,\lambda'}(-x)$ and the symmetry (\ref{appendix3.2}), taking the limit $\epsilon=0,\lambda=0$ and putting in the fermionic Green's functions we obtain for the sum of both terms with $\theta$ function
\begin{align}
 \text{Im} \Pi_{VC,\theta}(\mathbf Q, 2 \Delta + \delta)&= 2 \pi g^4 N \Delta \biggl( \frac{1}{8 \pi^2 v_\parallel v_\perp} \biggr)^2\cdot \delta \cdot   \int d\epsilon' \, d\lambda' \,  \text{Im} \chi_{0,\lambda'-\epsilon'}(0)   \sum_{\{A,B,C,D\}}   \times \nonumber \\
  & \quad    \mathcal P  \biggl( \frac{  v_{A,0}^2 u_{B,\epsilon'}^2   u_{C,0}^2u_{D,\lambda'}^2+  u_{A,\lambda'}^2  u_{B,0}^2 u_{C,\epsilon'}^2 v_{D,0}^2}{(\Delta+\delta-\xi_{\epsilon'})(-\Delta-\delta-\xi_{\lambda'})}  +  \frac{  v_{A,0}^2 v_{B,\epsilon'}^2   u_{C,0}^2u_{D,\lambda'}^2+  u_{A,\lambda'}^2  u_{B,0}^2 v_{C,\epsilon'}^2 v_{D,0}^2}{(\Delta+\delta+\xi_{\epsilon'})(-\Delta-\delta-\xi_{\lambda'})} \nonumber \\
& \quad  \quad  +   \frac{  v_{A,0}^2 u_{B,\epsilon'}^2   u_{C,0}^2 v_{D,\lambda'}^2+  v_{A,\lambda'}^2  u_{B,0}^2 u_{C,\epsilon'}^2 v_{D,0}^2}{(\Delta+\delta-\xi_{\epsilon'})(-\Delta-\delta+\xi_{\lambda'})} +   \frac{  v_{A,0}^2 v_{B,\epsilon'}^2   u_{C,0}^2 v_{D,\lambda'}^2+  v_{A,\lambda'}^2  u_{B,0}^2 v_{C,\epsilon'}^2 v_{D,0}^2}{(\Delta+\delta+\xi_{\epsilon'})(-\Delta-\delta+\xi_{\lambda'})}    \biggr)    \nonumber
  \end{align}  
With the tabular (\ref{appendix3.4}) we can perform the summation over the possible diagrams in (\ref{sumGF}) and find that for both $d$-wave and $s$-wave pairing the sum over the different combinations of the coherence factors are similar up to a sign
\begin{align*}
\sum_{\{A,B,C,D\}} \bigl(   v_{A,0}^2 u_{B,\epsilon'}^2   u_{C,0}^2u_{D,\lambda'}^2+  u_{A,\lambda'}^2  u_{B,0}^2 u_{C,\epsilon'}^2 v_{D,0}^2 \bigr) &= \frac{\epsilon' \lambda'}{2 \xi_{\epsilon'} \xi_{\lambda'}}  \, , \\
\sum_{\{A,B,C,D\}} \bigl(  v_{A,0}^2 v_{B,\epsilon'}^2   u_{C,0}^2u_{D,\lambda'}^2+  u_{A,\lambda'}^2  u_{B,0}^2 v_{C,\epsilon'}^2 v_{D,0}^2 \bigr) &=- \frac{\epsilon' \lambda'}{2 \xi_{\epsilon'} \xi_{\lambda'}}   \, ,\\
\sum_{\{A,B,C,D\}} \bigl(   v_{A,0}^2 u_{B,\epsilon'}^2   u_{C,0}^2 v_{D,\lambda'}^2+  v_{A,\lambda'}^2  u_{B,0}^2 u_{C,\epsilon'}^2 v_{D,0}^2 \bigr)&=- \frac{\epsilon' \lambda'}{2 \xi_{\epsilon'} \xi_{\lambda'}}   \, , \\
\sum_{\{A,B,C,D\}} \bigl(  v_{A,0}^2 v_{B,\epsilon'}^2   u_{C,0}^2 v_{D,\lambda'}^2+  v_{A,\lambda'}^2  u_{B,0}^2 v_{C,\epsilon'}^2 v_{D,0}^2 \bigr)&= \frac{\epsilon' \lambda'}{2 \xi_{\epsilon'} \xi_{\lambda'}} \, .
\end{align*}
This allow us to simplify the $\theta$-terms
\begin{align}
 \text{Im}  \Pi_{VC,\theta}(\mathbf Q, 2 \Delta + \delta)= 4 \pi  g^4  N \Delta  & \biggl( \frac{1}{8 \pi^2 v_\perp v_\parallel} \biggr)^2 \cdot \delta \cdot \int d\epsilon' \, d\lambda' \, \text{Im} \chi_{0,\lambda'-\epsilon'} (0) \, \times    \nonumber \\
&  \mathcal{P} \frac{\epsilon' \cdot \lambda'}{(\Delta+\delta+\xi_{\epsilon'})(\Delta+\delta+\xi_{\lambda'})(\Delta+\delta-\xi_{\epsilon'})(\Delta+\delta-\xi_{\lambda'})} \, .   \label{appendix3.a.4}
\end{align}
Obviously there will be no contributions to the discontinuity at $\omega = 2 \Delta$  if there were no divergences in the integral for $\delta \rightarrow 0$. Since the fraction of the integrand is an odd function and is  defined with the principal value, there is not singularity from the zeros of the denominator. Furthermore, the momentum dependence of the spin susceptibility leads to a convergence for large momentum transfers. Finally, we can conclude that there are no discontinuity-contributions for the imaginary part of the vertex correction by the $\theta$ terms.

\subsection*{Terms with $\delta$-function and two coherence factors}
With the help of the symmetry (\ref{appendix3.2}) it is always possible to rewrite the remaining terms in a way, that we have a $\delta$ function of the kind $\delta(\omega-\xi_{\mathbf k} - \xi_{\mathbf k + \mathbf Q})$ . Setting again $\omega = 2 \Delta + \delta$ the phase space will be restricted to $(\epsilon,\epsilon') \rightarrow (0,0)$ with:
\begin{align}
\int d\epsilon \, d\epsilon'\,  \delta( 2 \Delta + \delta - \xi_\epsilon - \xi_{\epsilon'}) f(\epsilon,\epsilon') \approx \int d\epsilon \, d\epsilon'\,  \delta( \delta- \frac{\epsilon^2+\epsilon'^2}{2\Delta}) f(\epsilon,\epsilon') \xrightarrow{\delta \rightarrow 0} 2 \pi \Delta f(0,0)   \, .  \label{appendix3.b.2}
\end{align}
Using this relation it follows that we can write the contribution to the discontinuity caused by the terms with the $\delta$-function and two coherence factors as:
\begin{align}
\delta_2 &= \lim_{\delta \rightarrow 0}\text{Im}  \Pi_{VC,\delta+2cf} (\mathbf Q, 2 \Delta + \delta)   \nonumber \\
&= 2 \pi \Delta g^4 N \biggl( \frac{1}{8 \pi^2 v_\perp v_\parallel} \biggr)^2 \int d\lambda \, d\lambda' \, \int_0^\infty dz \, \text{Im} \chi_{\lambda,\lambda'}(z) \times \nonumber \\
& \quad \mathcal P \biggl(   \frac{ f_{vv}(\lambda,\lambda') }{(-z+\Delta-\xi_{\lambda})( -z-\Delta-\xi_{\lambda'}  )}    +\frac{ f_{uv}(\lambda,\lambda')}{(-z+\Delta-\xi_{\lambda})( -z-\Delta+\xi_{\lambda'}  )}   \nonumber \\
& \quad  +\frac{f_{vu}(\lambda,\lambda')}{(-z+\Delta+\xi_{\lambda})( -z-\Delta-\xi_{\lambda'}  )}   +\frac{ f_{vv}(\lambda,\lambda')}{(-z+\Delta+\xi_{\lambda})( -z-\Delta+\xi_{\lambda'}  )}    \biggr)  \label{appendix3.b.3}
\end{align}
where we defined the $f$ functions, which can be calculated for the $d$-wave case as 
\begin{align}
\begin{split}
f_{uu}(\lambda,\lambda')&=\sum_{\{A,B,C,D\}} \bigl( v_{A,0}^2 u_{B,0}^2 u_{C,\lambda}^2 u_{D,\lambda'}^2   + u_{A,\lambda'}^2 u_{B,\lambda}^2 u_{C,0}^2 v_{D,0}^2   \bigr) =  \frac{\lambda \lambda'+(-\Delta+\xi_\lambda) (\Delta+\xi_{\lambda'})}{2 \xi_{\lambda} \xi_{\lambda'}}   \, ,  \\
f_{uv}(\lambda,\lambda')&=\sum_{\{A,B,C,D\}}   \bigl(v_{A,0}^2 u_{B,0}^2 u_{C,\lambda}^2 v_{D,\lambda'}^2+  v_{A,\lambda'}^2 u_{B,\lambda}^2 u_{C,0}^2 v_{D,0}^2 \bigr) =  \frac{-\lambda \lambda'+(-\Delta+\xi_\lambda) (-\Delta+\xi_{\lambda'})}{2 \xi_{\lambda} \xi_{\lambda'}}  \, ,   \\
 f_{vu}(\lambda,\lambda')&=\sum_{\{A,B,C,D\}}  \bigl(v_{A,0}^2 u_{B,0}^2 v_{C,\lambda}^2 u_{D,\lambda'}^2+  u_{A,\lambda'}^2 v_{B,\lambda}^2 u_{C,0}^2 v_{D,0}^2\bigr) =  \frac{-\lambda \lambda'+(\Delta+\xi_\lambda) (\Delta+\xi_{\lambda'})}{2 \xi_{\lambda} \xi_{\lambda'}}  \, , \\
 f_{vv}(\lambda,\lambda')&=\sum_{\{A,B,C,D\}}   \bigl( v_{A,0}^2 u_{B,0}^2 v_{C,\lambda}^2 v_{D,\lambda'}^2+  v_{A,\lambda'}^2 v_{B,\lambda}^2 u_{C,0}^2 v_{D,0}^2 \bigr)=  \frac{\lambda \lambda'+(\Delta+\xi_\lambda) (-\Delta+\xi_{\lambda'})}{2 \xi_{\lambda} \xi_{\lambda'}}  \, .
\end{split} \label{appendix3.b.4}
\end{align}
Fortunately, it is easy to show that these $f$ functions vanish in the $s$-wave case. As it will be shown in the following calculation all contributions to the discontinuity  from the vertex correction contain these $f$ functions and therefore we can immediately assess that also for the vertex corrected theory the phase sensitivity of the discontinuity is conserved. For this purpose we will restrict our further calculations to the $d$-wave case and using~(\ref{appendix3.b.4}) we find
\begin{align}
\delta_2&= 4 \pi \Delta g^4 N  \biggl( \frac{1}{8 \pi^2 v_\perp v_\parallel} \biggr)^2 \int d\lambda \, d\lambda' \, \int_0^\infty dz \, \text{Im} \chi_{\lambda,\lambda'}(z)  \mathcal P \frac{ \lambda \lambda' +z^2}{ \bigl[(\Delta-z)^2-\xi_\lambda \bigr]\bigl[(\Delta+z)^2-\xi_{\lambda'}^2 \bigr]} \label{appendix3.b.4.1}
\end{align}
Using (\ref{appendix3.b.2}), the symmetry $\chi_{\lambda,\lambda'}(x) =\chi_{\lambda,\lambda'}(-x)$ and the functions in (\ref{appendix3.b.4}) we easily obtain for the discontinuity contribution from the terms with the $\delta$ function and 3 or 4 coherence factors
\begin{align}
\delta_3 &= \lim_{\delta \rightarrow 0}\text{Im}  \Pi_{VC,\delta+3cf} (\mathbf Q, 2 \Delta + \delta)   \nonumber \\
&=  2 \pi^2 \Delta g^4 N  \biggl( \frac{1}{8 \pi^2 v_\perp v_\parallel} \biggr)^2 \int  d\lambda \, d\lambda' \,  \biggl[ \chi_{\lambda,\lambda'}( \Delta+\xi_{\lambda} ) \,   \mathcal P \biggr(  \frac{ f_{vu}(\lambda,\lambda')}{-2 \Delta-\xi_{\lambda}-\xi_{\lambda'}}    +  \frac{ f_{vv}(\lambda,\lambda')}{-2 \Delta-\xi_{\lambda}+\xi_{\lambda'}}  \biggr)   \nonumber \\ 
& \hspace*{5.5cm}  +  \chi_{\lambda,\lambda'}(\xi_{\lambda'}-\Delta) \, \mathcal P \biggl( \frac{f_{uv}(\lambda,\lambda')}{2 \Delta-\xi_{\lambda'}-\xi_\lambda} + \frac{f_{vv}(\lambda,\lambda')}{2 \Delta-\xi_{\lambda'}+\xi_\lambda} \biggr)   \biggr]     \label{appendix3.c.1}   \\
\delta_4 &= \lim_{\delta \rightarrow 0} \text{Im}  \Pi_{VC,\delta+4cf} (\mathbf Q, 2 \Delta + \delta)   \nonumber \\
&=- 2 \pi^3 \Delta g^4 N \biggl( \frac{1}{8 \pi^2 v_\perp v_\parallel} \biggr)^2 \int d\lambda d\lambda' \text{Im}  \chi_{\lambda,\lambda'}(\Delta + \xi_\lambda) \delta( 2 \Delta + \xi_\lambda - \xi_{\lambda'}) f_{vv}(\lambda,\lambda')     \label{appendix3.d}
\end{align}
At this point of the calculation it will be useful to apply the approximation that the parallel and perpendicular Fermi velocities are equal, see Eq.~(\ref{appendix1.6}). Thus, the bosonic propagator will be an even function in $\lambda$ and $\lambda'$ and odd terms in the $f$ functions will vanish due to the antisymmetry of the complete integrand. This allows us to simplify
\begin{align}
\begin{split}
f_{uu}(\lambda,\lambda')&= \frac{(-\Delta+\xi_\lambda) (\Delta+\xi_{\lambda'})}{2 \xi_{\lambda} \xi_{\lambda'}}   \, ,  \\
f_{uv}(\lambda,\lambda')&=  \frac{(-\Delta+\xi_\lambda) (-\Delta+\xi_{\lambda'})}{2 \xi_{\lambda} \xi_{\lambda'}}  \, ,   \\
 f_{vu}(\lambda,\lambda')&=\frac{(\Delta+\xi_\lambda) (\Delta+\xi_{\lambda'})}{2 \xi_{\lambda} \xi_{\lambda'}}  \, , \\
 f_{vv}(\lambda,\lambda')&= \frac{(\Delta+\xi_\lambda) (-\Delta+\xi_{\lambda'})}{2 \xi_{\lambda} \xi_{\lambda'}}  \, .
\end{split} \label{appendix3.d.3}
\end{align}
We now introduce dimensionless integration variables $x=\frac{\epsilon}{\Delta}, y= \frac{\epsilon'}\Delta, \nu= \frac z \Delta$, the dimensionless dispersion $\tilde \xi_x = \sqrt{x^2+1}$ and the dimensionless RPA spin susceptibility
\begin{align}
\tilde \chi_{x,y}(\nu) &= \gamma \Delta \chi_{x \cdot \Delta, y \cdot \Delta} (\nu \cdot \Delta) = \gamma \Delta  \frac{1}{r+ c_s \frac{(x \cdot \Delta )^2+(y \cdot\Delta)^2}{v_F^2} - \Pi_{\mathbf Q}^{(2)}(\nu \cdot \Delta)} = \frac{1}{\frac{\omega_{\text{sf}}}\Delta+ \hat \Delta  (x^2+y^2) -\frac{\Pi_{\mathbf Q}^{(2)}(\nu \cdot \Delta)}{\gamma \Delta}}    \, , \label{appendix3.d.4}
\end{align}
which is a similar definition as in Eq.~(\ref{ase4}). Analogously, we defined the dimensionless pairing parameter $\hat \Delta = \frac{c_s \Delta}{v_F^2 \gamma}$. Substituting these variables in the above formulas for $\delta_1$, $\delta_2$ and $\delta_3$, evaluating the remaining $\delta$-function and simplifying the expressions leads to
\begin{align}
\delta D =& \delta_2+\delta_3+\delta_4  =  \frac{D_0}{N}  \cdot \kappa(\frac{\omega_{\text{sf}}}\Delta, \hat \Delta)\label{appendix3.d.5}
\end{align}
with the dimensionless function
\begin{align}
\kappa(\frac{\omega_{\text{sf}}}\Delta, \hat \Delta) =&  \frac{4}{\pi^2}  \int_0^\infty dx \, dx \, d\nu \, \text{Im}  \tilde \chi_{x,y}(\nu)  \mathcal P \frac{ \nu^2}{ \bigl[(1-\nu)^2-\tilde \xi_x^2 \bigr]\bigl[(1+\nu)^2-\tilde \xi_y^2 \bigr]}     \nonumber  \\
&  +\frac{2}{\pi} \int_0^\infty  dx \, dy \,   \mathcal P \biggl[ \frac{\tilde \chi_{x,y}( 1+\tilde \xi_x) }{\tilde \xi_x} \frac{(1+\tilde \xi_x)^2}{\tilde \xi_y^2 - (2 +\tilde \xi_x)^2} +\frac{\chi_{x,y}( \tilde \xi_y-1 ) }{\tilde \xi_y} \frac{ (1-\tilde \xi_y)^2}{\tilde \xi_x^2 - (2 -\tilde \xi_y)^2}  \biggr]   \nonumber \\
& -   \int_0^\infty dx \, \text{Im}  \tilde \chi_{x,\sqrt{(2+\tilde \xi_x)^2-1}}(1+ \tilde \xi_x) \, \frac{(1+ \tilde \xi_x)^2}{\tilde \xi_x \sqrt{(2+\tilde \xi_x)^2-1}}  \, .   \label{appendix3.d.6}
\end{align}

 \subsection*{Numerical analysis of the vertex correction of discontinuity}    \label{appendixnumerical}
At this point of the calculation we need a good input for the 1-loop spin susceptibility in the superconducting state in order to estimate the numerical function $\kappa(r', \hat \Delta)$. From~(\ref{appendix3.d.6}) we see that both the resonance region $\omega < 2 \Delta$ and the continuum region $\omega>2 \Delta$ contribute to the discontinuity correction. In the continuum region $\omega > 2 \Delta$ we can approximate the self-energy to be similar to the normal state polarization operator $ \Pi_{\mathbf Q}(\omega) \approx - i \gamma \omega \, \theta(\omega-2 \Delta)$ with an additional spin gap below $2 \Delta$, because for $\omega > \Delta$ the superconducting and the normal conducting properties are quite similar. In the resonance region $\omega<2 \Delta$ we use the known results from the one-loop calculations and estimate the dimensionless spin susceptibility to be 
\begin{align}
\tilde \chi_{x,y}(\nu<2) &= - \frac{1}{\pi} \frac{\tilde Z_{\text{res}}}{\nu- \tilde \Omega_{\text{res}}+i0}   \qquad \qquad \text{with } \tilde Z_{\text{res}} = 2 \pi  \biggl(1- \frac{\tilde \Omega_{\text{res}}}{2}   \biggr)   \, ,   \label{appendix3.d.7}
\end{align}
where dimensionless resonance energy is defined as $\tilde \Omega_{\text{res}} = 2 \bigl(1- e^{- \bigl[\frac{\omega_{\text{sf}}}{\Delta} + \hat \Delta (x^2+y^2)   \bigr]} \bigr)$. The imaginary part of the $\tilde \chi_{x,y}$ is then just given by Eq.~(\ref{ase4}). As already stressed the two external parameters $\frac{\omega_{\text{sf}}}{\Delta}$ and $\hat \Delta$ can be expressed in terms of $C_2(\lambda)$ and $\lambda$ using Eq.~(\ref{ase7}). Using these approximations it is possible to determine numerically the function $\kappa(\frac{\omega_{\text{sf}}}\Delta, \hat \Delta) = \tilde \kappa(\lambda)$. 

In the limit of large $\lambda \gg 1$ the resonance energy $\Omega_{\text{res}} \ll \Delta$ at  $\mathbf{q} =\mathbf{Q}$. For small $x,y \ll \hat \Delta^{-1} \sim 1$, this will lead to a different dependence of the resonance energy $\Omega_{\text{res}} \sim \sqrt{\omega_{\text{sf}}/{\Delta} + \hat \Delta (x^2+y^2)} $ and of the spectral weight $\tilde Z_{\text{res}} \sim \bigl(\sqrt{\omega_{\text{sf}}/{\Delta} + \hat \Delta (x^2+y^2)}  \bigr) ^{-1}$. For $\lambda \gg 1$, $\omega_{\text{sf}} \ll {\Delta}$ and consequently the spectral weight for small $x,y \ll \hat \Delta^{-1}$ will be larger than for the exponential dependence in Eq.~(\ref{appendix3.d.7}). Nevertheless, as this different behavior is restricted to a small $x,y$ phase space we do not expect qualitative changes of the occuring integrals in~(\ref{appendix3.d.6}).

\section{Phase sensitivity at three-loop}   \label{Appendix3loop}
Let us consider the possible three-loop diagrams without the diagrams containing the one-loop self-energy correction that was already calculated in Eq.~(\ref{secfeyn}), see Fig.~\ref{3loop}. Note that we have no diagrams containing a bosonic line with particle-hole bubble since they are already included in the self-consistent two-loop diagram~(\ref{smpapervc3}).  We want to show that the discontinuity for the s-wave symmetry $\Delta_{\mathbf{k_F+Q}}=\Delta_{\mathbf{k_F}}$ vanished also for these diagrams. As can be seen in the previous Appendix~\ref{appendix3} the restriction to external frequency $\omega=2 \Delta$ and momenta $\mathbf Q$ constrain two fermionic propagators  connected to the left or right external boson line to lie directly at the hot spots. In the two-loop calculation this allowed us to combine all important information about the coherence factors and different combinations of normal and anomalous Green's functions in the functions defined in Eq.~(\ref{appendix3.b.4}). From now on we will always restrict  the two propagators on the left side to the hot spots. The first diagram in Fig.~\ref{3loop} is given by
\begin{align}
\Pi^{(3)}_{(a)} &\sim \sum_{i,j} \sum_{k,q,q'} \chi_q \chi_{q'} \text{tr} \bigl[ \hat {\mathcal{G}}_{k} \alpha^z \hat {\mathcal{G}}_{k+Q} \alpha^i \hat {\mathcal{G}}_{k+Q+q} \alpha^j  {\mathcal{G}}_{k+Q+q+q'} \alpha^z {\mathcal{G}}_{k+q+q'} \alpha^j {\mathcal{G}}_{k+q} \alpha^j  \bigr]   \nonumber \\
&= \sum_{k,q,q'} \sum_{\{A,B,C,D,E,F\}} \chi_q \chi_{q'}   A_{k} B_{k+Q} C_{k+Q+q} D_{k+Q+q+q'} E_{k+q+q'} F_{k+q}    \label{3la}
\end{align}
\begin{figure}
\centering
\includegraphics[width=0.77\textwidth]{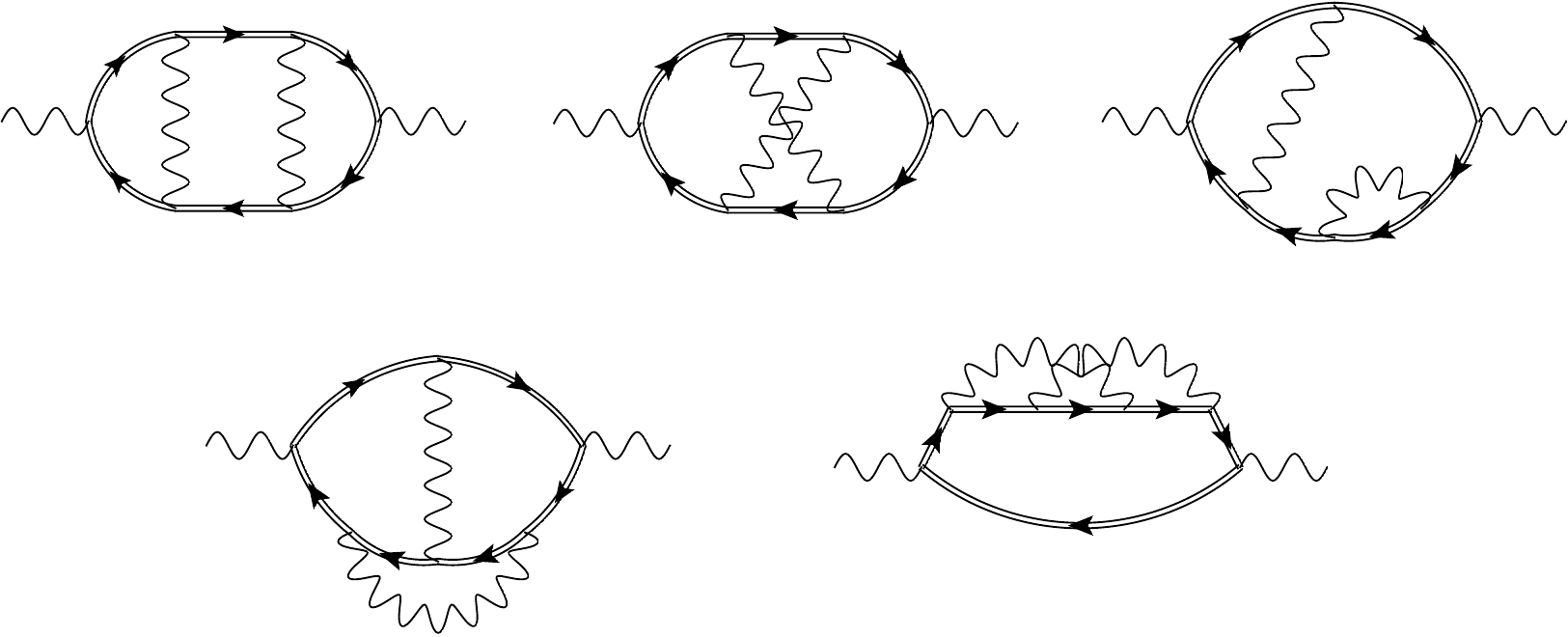}
\caption{Three-loop diagrams without one-loop self-energy contributions}
\label{3loop}
\end{figure}

\noindent where $Q=(\omega,\mathbf Q)$ is the external frequency and momentum. In the second step we performed the trace over the Nambu and spin degrees of freedom and  found again all possible diagrams $\{A,B,C,D,E,F\}$ with arrow conservation at each vertex. Assuming that only the scattering between fermions around the hot spots contribute (so $\chi_q$ is strongly peaked around $\mathbf Q$) we are able to linearize the dispersions.  Thus we get for the discontinuity contribution from the diagram (a) 
\begin{align}
\delta D^{(3)}_{(a)} \sim  \int d \epsilon_1 \, d\epsilon_2 \, d\epsilon_3 \, d\epsilon_4  \, \chi_1(\{\epsilon_i\}) \,  \chi_2(\{\epsilon_i\}) \sum_{\{A,B,C,D,E,F\}}  v_{A,0}^2 u_{B,0}^2 u_{C,\epsilon_1}^2 u_{D,\epsilon_2}^2 u_{E,\epsilon_3}^2 u_{F,\epsilon_4}^2 + \ldots \, .  \label{3lb}
\end{align}
where we set the two left propagators $A$ and $B$ at the hot spots and . There are several other contributions containing different combinations of the coherence factors of $C,D,E,F$ and we linearized the momenta in such a way that the fermionic propagators have independent momentum integration variables $\epsilon_1, \ldots, \epsilon_4$. The bosonic modes now contain the information about the complicated dependence of the different momentum integrations. Nevertheless, the explicit form of this dependence is not of importance, because the sum over all possible diagrams $\{A,B,C,D,E,F\}$, which can be obtained by evaluating the trace in~(\ref{3la}), vanishes for the s-wave case as can be shown by  coherence factors similar to~(\ref{appendix3.4}) and~(\ref{appendix3.5}). This behavior is not depending on the combination of coherence factors we wrote in~(\ref{3lb}) and performing the Nambu and spin traces over the remaining diagrams in Fig.~\ref{3loop} we can show that the combination of Green's functions $\{A,B,C,D,E,F\}$ are always the same, except of an overall factor. We also calculated several four-loop diagrams in a similar manner and found the same result. Summarizing, we could show that the phase sensitivity of the discontinuity of the bosonic self-energy is conserved up to three-loop order  and presented a procedure to systematically examine the behavior of the discontinuity depending on the superconducting gap symmetry in arbitrary order perturbation theory. 

\section{$\varepsilon$ expansion for vertex}   \label{appendixepsilon}
The most straightforward way to see the suppression of the vertex correction in higher powers of $g$ in  $d =3-\varepsilon$ dimensions for the strong coupling case $r \ll \gamma \Delta$ is to look at the vertex
\begin{align}
\delta \Gamma_{k,Q} &= \begin{matrix}
\includegraphics[width=3.4cm]{deltaGamma2}
\end{matrix}
\end{align}
with external momenta at the hot spots $\mathbf k = \mathbf k_F$ and the magnetic ordering vector $\mathbf Q$ on the imaginary axis. We will only consider the $G ^{(p)} G ^{(p)}$ combination of Green's functions for zero temperatures and at the phase transition $r=0$, resulting in
\begin{align}
 \delta \Gamma_{\mathbf k_F,\mathbf Q}(i \Omega, i \omega) &\sim g^3 \int d^{3-\varepsilon} \mathbf q \int d\nu \frac{1}{c_s (\mathbf q-\mathbf Q)^2 -\Pi_{\mathbf q}(i \nu)} \frac{i (\Omega +\nu)+  \varepsilon_{\mathbf k_F+\mathbf q}}{(\Omega+\nu)^2+\xi_{\mathbf k_F+\mathbf q}^2} \frac{i (\Omega+\omega +\nu)+  \varepsilon_{\mathbf k_F+\mathbf q+\mathbf Q}}{(\Omega+\omega+\nu)^2+\xi_{\mathbf k_F+\mathbf q+\mathbf Q}^2}   \nonumber \\
 &\sim g^3 \int dq_z^{1-\varepsilon} d \lambda d\lambda' d\nu \, \frac{1}{c_s q_z^2+\frac{ c_s(\lambda^2+\lambda'^2)}{v_F^2} - \Pi_{\mathbf Q}(i \nu)} \frac{i (\Omega +\nu)+  \lambda}{(\Omega+\nu)^2+\lambda^2+\Delta^2} \frac{i (\Omega+\omega +\nu)+  \lambda'}{(\Omega+\omega+\nu)^2+\lambda'^2+\Delta^2}   \, .
 \end{align} 
 Here, we assumed that only the bosonic propagator has a momentum dependence in the additional dimension $q_z$ and linearized the remaining two dimensional $(q_x,q_y)$ integration in the usual way. The first integration can be performed by $dq_z^{1-\varepsilon} = \Omega_{1-\varepsilon} dq_z z^{-\varepsilon}$, where $\Omega_d = 2 \pi^{d/2} / \Gamma(d/2)$ is the solid angle  in $d$ dimensions. A renormalization of the energy scales
 $$ x=\frac{\lambda}{\Delta}, y = \frac{\lambda'}{\Delta}, \tilde \nu = \frac{\nu}{\Delta} , \tilde \omega = \frac{\omega}{\Delta}, \tilde \Omega = \frac{\Omega}{\Delta}$$
yields the vertex
 \begin{align}
\delta \Gamma_{\mathbf k_F,\mathbf Q}(i \Omega, i \omega) &\sim \frac{g^3 \Delta^{\frac{1-\varepsilon}{2}}}{\gamma^{\frac{1+\varepsilon}{2}}} \int  dx dy d \tilde \nu \, \biggl(\frac{1}{\hat \Delta (x^2+y^2) - \frac{\Pi_{\mathbf Q}(i \tilde \nu \Delta)}{\gamma \Delta}}  \biggr)^{\frac{1+\varepsilon}{2}} \frac{i (\tilde \Omega + \tilde \nu)+ x}{(\tilde \Omega+\tilde \nu)^2+x^2+1} \frac{i (\tilde \Omega+\tilde \omega +\tilde \nu)+  y}{(\tilde \Omega+\tilde \omega+\tilde \nu)^2+y^2+1}  \nonumber \\
&\sim  g^{\frac{2}{\varepsilon}-1}  \int  dx dy d \tilde \nu \, \biggl(\frac{1}{\hat \Delta (x^2+y^2) - \frac{\Pi_{\mathbf Q}(i \tilde \nu \Delta)}{\gamma \Delta}}  \biggr)^{\frac{1+\varepsilon}{2}} \frac{i (\tilde \Omega + \tilde \nu)+ x}{(\tilde \Omega+\tilde \nu)^2+x^2+1} \frac{i (\tilde \Omega+\tilde \omega +\tilde \nu)+  y}{(\tilde \Omega+\tilde \omega+\tilde \nu)^2+y^2+1} \, .    \label{dG4}
 \end{align}
In the last step we used $\gamma \sim g^2$ and the result from the pairing instability $\Delta \sim g^{\frac{4-2\varepsilon}{\varepsilon}}$. The bosonic self-energy can be approximated on the imaginary axis as
 \begin{align}
 \frac{\Pi_ {\mathbf Q}(i \tilde \nu \Delta)}{\gamma \Delta} &\approx  \begin{cases} - \tilde \nu^2 & \text {for } \tilde \nu < 2    \\ -  |\tilde \nu| & \text {for } \tilde \nu > 2  \end{cases}
 \end{align}
 and is therefore just a function not depending on $g$. The only $g$-dependence in the integrand of~(\ref{dG4}) is hidden in the parameter $\hat \Delta \sim \frac{\Delta}{\gamma} \sim g^{\frac{4-4\varepsilon}{\varepsilon}}$. There can be critical contributions where
\begin{align}
\biggl(\frac{1}{\hat \Delta (x^2+y^2) - \frac{\Pi_{\mathbf Q}(i \tilde \nu \Delta)}{\gamma \Delta}}  \biggr)^{\frac{1+\varepsilon}{2}} \sim \biggl(\frac{1}{\hat \Delta}  \biggr)^{\frac{1+\varepsilon}{2}}    \label{crit1}
\end{align}  
for the resonance and continuum region, which can be separated from the non-critical terms
\begin{align}
\delta \Gamma_{\mathbf k_F,\mathbf Q}(i \Omega, i \omega) &= a(i \Omega, i \omega) \cdot g^{\frac{2}{\varepsilon}-1} + \delta \Gamma_{\mathbf k_F,\mathbf Q}^{\text{crit,res}}(i \Omega, i \omega)+ \delta \Gamma_{\mathbf k_F,\mathbf Q}^{\text{crit,con}}(i \Omega, i \omega)
\end{align}
Here, the dimensionless function $a(i \Omega, i \omega)$ is not $g$ dependent. These terms are for $\varepsilon<1$ not critical, because the vertex correction $\frac{\delta \Gamma}{\Gamma_0}  \sim g^{\frac{2}{\varepsilon}-2}$ is suppressed by a higher power in $g$ compared to the bare vertex $\Gamma_0 = g$. The question is how to estimate the critical contributions. For the resonance region $\tilde \nu<2$ the condition~(\ref{crit1}) is fulfilled for 
\begin{align}
 \hat \Delta (x^2+y^2) > \frac{\Pi_{\mathbf Q}(i \tilde \nu \Delta)}{\gamma \Delta} \approx \tilde \nu^2  |\tilde \nu| < \sqrt{\hat \Delta} \sim g^{\frac{2-2\varepsilon}{\varepsilon}}
 \end{align} 
and gives an exponentially small integration area for the $\tilde \nu$ integration. Thus the most critical contributions from the above integral~(\ref{dG4}) of the resonance region are
\begin{align}
\delta \Gamma_{\mathbf k_F,\mathbf Q}^{\text{crit,res}}(i \Omega, i \omega) \sim g^{\frac{2}{\varepsilon}-1} \int_{-\sqrt{\hat \Delta}}^{\sqrt{\hat \Delta}}d \tilde \nu \biggl(\frac{1}{\hat \Delta}  \biggr)^{\frac{1+\varepsilon}{2} } \sim g^{\frac{2}{\varepsilon}-1} \hat \Delta^{-\frac{\varepsilon}2} \sim g^{\frac{2}{\varepsilon}+2 \varepsilon-3} \, .
\end{align}
In an analogue treatment for the continuum region the critical contributions can analogue be estimated to at least of order 
\begin{align}
\delta \Gamma_{\mathbf k_F,\mathbf Q}^{\text{crit,con}}(i \Omega, i \omega) \sim g^{\frac{4}{\varepsilon}+ 2 \varepsilon- 5}  \, .
\end{align}
Therefore, for small $\varepsilon<1$ the so-called critical contributions of the vertex corrections
$$\frac{\delta \Gamma_{\mathbf k_F,\mathbf Q}^{\text{crit,res}}}g  \sim g^{\alpha_1(\varepsilon)}    \qquad \qquad \frac{\delta \Gamma_{\mathbf k_F,\mathbf Q}^{\text{crit,con}}}g    \sim g^{\alpha_2(\varepsilon)} \qquad \qquad \text{with } \alpha_1(\varepsilon),\alpha_2(\varepsilon) > 0$$
 are suppressed by a higher order in $g$ and therefore negligible. 

\end{widetext}

\end{document}